\documentclass[%
 aip,
 amsmath,amssymb,
 reprint,%
]{revtex4-1}

\usepackage{graphicx}
\usepackage{dcolumn}
\usepackage{bm}

\usepackage[utf8]{inputenc}
\usepackage[T1]{fontenc}
\usepackage{mathptmx}
\usepackage{etoolbox}
\usepackage{bbding}
\makeatletter
\def\@email#1#2{%
 \endgroup
 \patchcmd{\titleblock@produce}
  {\frontmatter@RRAPformat}
  {\frontmatter@RRAPformat{\produce@RRAP{*#1\href{mailto:#2}{#2}}}\frontmatter@RRAPformat}
  {}{}
}%
\usepackage{subfig,float,epstopdf,epsfig,url,hyperref,diagbox,xcolor}
\usepackage{mathtools}
\DeclarePairedDelimiter\abs{\lvert}{\rvert}%
\DeclarePairedDelimiter\norm{\lVert}{\rVert}%
\makeatletter
\let\oldabs\abs
\def\abs{\@ifstar{\oldabs}{\oldabs*}}
\let\oldnorm\norm
\def\norm{\@ifstar{\oldnorm}{\oldnorm*}}
\usepackage[makeroom]{cancel}
\makeatother
\begin{document}

\title{An exposition on some viscosity dependent issues in refined potential flow theory}
\author{Taofiq Omoniyi Amoloye}

\affiliation{Department of Aeronautical and Astronautical Engineering,
Faculty of Engineering and Technology, Kwara State University, Malete, Nigeria.}
\author{Leke Thaddeus Oladimeji}

\affiliation{Department of Aeronautical and Astronautical Engineering,
Faculty of Engineering and Technology, Kwara State University, Malete, Nigeria.}
\author{Mahmoud A. Hayajnh}
	
\affiliation{Department of Aeronautical Engineering,
Jordan University of Science and Technology, Irbid, Jordan 22110.}
\author{Olalekan Adebayo Olayemi}
\affiliation{Department of Aeronautical and Astronautical Engineering,
Faculty of Engineering and Technology, Kwara State University, Malete, Nigeria.}
\author{Abdulfatai Olanrewaju Sulaiman}
\affiliation{Department of Aeronautical and Astronautical Engineering,
Faculty of Engineering and Technology, Kwara State University, Malete, Nigeria.}

\email{abdulfatai.sulaimano@kwasu.edu.ng}
\email{olalekan.olayemi@kwasu.edu.ng}
\email{mahayajnh@just.edu.jo}
\email{leke.oladimeji14@gmail.com}
\email{taofiq.amoloye@kwasu.edu.ng}

\date{\today}

\begin{abstract}
This study explores viscosity-dependent issues in refined potential flow theory (RPT) applied to the incompressible flow over an impulsively started circular cylinder, a classical fluid dynamics problem with significant engineering applications. RPT builds on previous theoretical, numerical, and experimental studies, refining classical potential flow theory to incorporate viscous effects, flow fluctuations, and three-dimensional influences. The governing equations, including the continuity and Navier-Stokes equations, are solved using a quasi-potential viscous stream function that satisfies the free-stream and no-slip boundary conditions. The current study investigates wake characteristics at various Reynolds numbers ($Re$) and non-dimensional times, focusing on phenomena such as forewake formation, vortex shedding, and turbulence decay. The results show that RPT accurately predicts certain features, including Strouhal numbers and turbulence decay rates, which align with experimental and computational fluid dynamics data. However, discrepancies remain in modeling the initial transient flow development and flow behavior across wide $Re$ range. The analysis highlights the critical role of viscosity and cylinder radius in influencing flow stability, symmetry, and energy spectra. Lower viscosity promotes smaller-scale structures and higher-frequency oscillations, while higher viscosity redistributes energy to lower frequencies and larger-scale features. The study concludes that, while RPT provides valuable insights into wake dynamics, further refinement is needed to address its limitations, particularly in modeling viscosity-dependent phenomena. These findings contribute to the advancement of the refined potential flow theory and its application in fluid dynamics and engineering design.
\end{abstract}

\maketitle

\section{\label{Intro}Introduction}
Incompressible flow over a circular cylinder is a classical fluid dynamics problem. The study and understanding of the dynamics of the flow is important due to its wide application in engineering design and configurations. These studies cover theoretical \cite{Aboelkassem2026,Amoloye2024,UedaandKida2021, MatheswaranandMiller2024a, MatheswaranandMiller2024b,BarlevandYang1975}, numerical \cite{Chatzietal2022,KoumoutsakosandLeonard1995}, and experimental \cite{BouardandCout1980,Coutanceauetal1985,CoutandBouard1977a,CoutandBouard1977b} work from the foundational research \cite{Strouhal1878,Rayleigh1879,Weisfred2017,Tollmienetal1961} to the present day. The governing equations are the continuity and the Navier-Stokes equations. The persistent difficulty of a theoretical analysis of the problem is propagated from early studies of d'Alembert's paradox and is emblematic of the Navier-Stokes problem\cite{Amoloye2024,CMI}. 

Amoloye \cite{Amoloye2024,Amoloyeetal2026b} refined the classical potential flow theory to model the unsteady, viscous, incompressible, and three-dimensional flow, focusing on the wake characteristics of an impulsively started circular cylinder in the sub-critical regime. The refinement utilizes experimental and CFD observations of the cylinder flow. Important $Re$-dependent features of the wake are captured, including excellent predictions of the shedding frequency and the inertial sub-range of turbulence,, particularly at $Re=3,900$ \cite{Amoloye2024} (based on the free stream velocity, $V_\infty$, the cylinder diameter, $D$, and the fluid kinematic viscosity, $\nu$, as $Re=V_{\infty}D/\nu$). However, the model deviates from the experimental and computational fluid dynamics (CFD) results on a number of viscosity-dependent issues, including the flow rapidity, the velocity in the core of the wake eddies and on the rear axis of the cylinder \color{black}(outlier points \cite{Amoloye2024}), \color{black}and wake characteristics in the laminar regime \cite{Amoloye2024}.

Therefore, the present publication seeks to explore some of these issues. Section~\ref{GEBC} discusses the governing equations and boundary conditions. Section~\ref{RPT} gives an overview of the refined potential flow theory (RPT). The equi-vorticity contours, instantaneous streamline patterns, and spatial velocity spectra are reported and discussed compared to published results from previous RPT analysis, experiments and CFD in Section~\ref{RnD}. Section~\ref{C} concludes the paper.

\section{\label{GEBC}Governing Equations and Boundary Conditions}
The governing equations for the incompressible cylinder crossflow of a Newtonian fluid \color{black}with constant properties \color{black}considered here are the continuity equation 
\begin{equation}
\label{contdiff}
\nabla{}\cdot\mathbf{V}=0\hspace{10pt} (t\ge 0),
\end{equation}
and the Navier-Stokes equations (NSE)
\begin{equation}
\label{NSEDiff}
\begin{array}{l}
\dfrac{\partial{}\mathbf{V}}{\partial{}t}+\nabla{}\left(\dfrac{p}{\rho{}}+\dfrac{V^2}{2}-\left[\dfrac{\lambda}{\rho}+2\nu{}\right]\nabla{}\cdot \mathbf{V}\right)\\[10pt]=\mathbf{V}\times{}\omega{}-\color{black}\nu{}\nabla{}\times{}\omega{}\color{black}
\hspace{10pt} (t\ge 0)
\end{array}
\end{equation}
where $\mathbf{V}$ is the velocity, $p$ is the pressure, $\rho$ is the density, $\omega$ is the vorticity vector, $\lambda$ is the second coefficient of viscosity, $\nu$ is the kinematic viscosity, and $t$ is the dimensional time \cite{Anderson2011}. Together, Eqs.~\ref{contdiff} and \ref{NSEDiff} constitute the Navier-Stokes problem \cite{CMI,Farwig2021} that is an initial boundary value problem. The boundary conditions are the infinity boundary condition at the free stream and the no-slip boundary condition at the cylinder surface.

\section{An Overview of Refined Potential Flow Theory}\label{RPT}
The domain for the cylinder flow is described by the three-dimensional Cartesian geometric coordinate axes as
\begin{equation}
\label{3Ddomain}
\begin{array}{l}
\hspace{10pt}x=r\cos{\theta}\sin{{\varphi}} \hspace{10pt}y=r\sin{\theta}\sin{{\varphi}} \hspace{10pt}z=r\cos{{\varphi}}\hspace{3pt}.
\end{array}
\end{equation}
$r$, $\theta$, and $\varphi$ are the radial, tangential, and azimuthal coordinate variables of the affiliated spherical geometric coordinate system, respectively. Attention is briefly focused on the azimuthal plane where $\varphi=\pi/2$ or $z=0$. To allow a nonzero $\omega$ and automatically satisfy Eq.~\ref{contdiff}, $\mathbf{V}$ is described by a vector of two-dimensional stream functions as
\begin{equation}
\label{SFVect}
\mathbf{V}=\nabla{}\times\overrightarrow{\psi{}},
\end{equation}
where $\overrightarrow{\psi}=\psi_{vis}\hat{\mathbf{e}}_z$\cite{Amoloye2024,Amoloyeetal2026b} \color{black}(vis: viscous) \color{black} and other components of the vector are zero. RPT prescribes the unsteady viscous stream function, $\psi_{vis}$, as a superposition of individual stream functions, that is, 

\begin{equation}
\label{NewPsiVis}
\psi_{vis}={\psi{}}_{inv}+{\psi{}}_{vt}+{\psi{}}_{ss}+{\psi{}}_{p}+{\psi{}}_{vtsd}  \hspace{3pt}	
\end{equation}
(vis: viscous, inv: inviscid, vt: vortex sheet, ss: sink/source sheet, p: perifocal, and vtsd: vortex shedding), each modeling different features of the cylinder flow where $R$ is the radius of the cylinder \cite {Amoloye2020,Amoloye2024}. Here, Eqs.~\ref{NewPsiInv}-\ref{NewPsiVtsd} describe each of the constituent functions of Eq.~\ref{NewPsiVis}, respectively, without derivation. 
\begin{equation}
	\label{NewPsiInv}
	\begin{array}{l}
	{\psi{}}_{inv} = V_{\infty{}}r\sin{\theta{}}\left(1-\frac{R^2}{r^2}\right)\exp{\left[-\frac{R^2}{4V^2_\infty t^2} \right]}
				\hspace{3pt},
	\end{array}	
	\end{equation}
\begin{widetext}
\begin{equation}
	\label{NewPsiVt}
	\begin{array}{l}
	{\psi{}}_{vt}=-V_{\infty}r\sin{\theta}\left(1+\frac{R^2}{r^2}\right) {\exp{\left[-\left(\frac{r^{2}}{4V_{\infty}^{2}t^{2}}+\frac{qRe\sqrt { \left( 1-{\frac {{R}^{2}}{{r}^{2
					}}} \right) ^{2}+4\,{\frac {{R}^{2} \sin^2{\theta} }{{r}^{2}}}+2}\left( {\frac {r}{R}}-1 \right)}{8} \right) \right]}}\ln{\left(\dfrac{r}{R}\right)}
				\hspace{3pt},
	\end{array}	
	\end{equation}
\end{widetext}
\begin{widetext}
\begin{equation}
	\label{NewPsiSS}
	\begin{array}{l}
	{\psi{}}_{ss}=V_{\infty{}}r\left(1-\frac{R^2}{r^2}\right)\left(\theta{}+\sigma{}\right)\left(\chi\sin{\theta{}}\right)^{\exp{\left[-\frac{\sqrt{\nu t}}{r}\left(1-\frac{R}{r}\right)\right]}}\\[10pt]\times\left(\cos{\theta{}}-\frac{qRe}{2}\, {\frac {\sin^2{\theta }\cos{\theta} \left( {\frac {
					r}{R}}-1 \right)\frac{{R}^{2}}{{r}^{2}} }{\sqrt {
				\left( 1-{\frac {{R}^{2}}{{r}^{2}}} \right) ^{2}+4\,{\frac {{R}^{2}
						\sin^2{\theta}}{{r}^{2}}}+2}}}\right) {\exp{\left[-\left(\frac{r^{2}}{4V_{\infty}^{2}t^{2}}+\frac{qRe\sqrt { \left( 1-{\frac {{R}^{2}}{{r}^{2
					}}} \right) ^{2}+4\,{\frac {{R}^{2} \sin^2{\theta} }{{r}^{2}}}+2} \left( {\frac {r}{R}}-1 \right)}{8}\right) \right]}}\ln{\left(\dfrac{r}{R}\right)}
				\hspace{3pt},
	\end{array}	
	\end{equation}
\end{widetext}
\begin{widetext}
\begin{equation}
	\label{CylGravityVelPot}
	\psi_{p}= -\dfrac{2\pi{\mu}^2_{g}}{r}\dfrac{\partial}{\partial \theta}\left(\left[\dfrac{2{\mu}_{g}}{r}-\begin{array}{l}\left(\dfrac{1}{r}\left[\dfrac{\partial\psi_{inv}}{\partial\theta}+\dfrac{\partial\psi_{vt}}{\partial\theta}+\dfrac{\partial\psi_{ss}}{\partial\theta}\right]\right)^2 +\left(-\left[\dfrac{\partial\psi_{inv}}{\partial r}+\dfrac{\partial\psi_{vt}}{\partial r}+\dfrac{\partial\psi_{ss}}{\partial r}\right]\right)^2\end{array}\right]^{-3/2}\right)
	\hspace{3pt},
\end{equation}
\end{widetext}
\begin{widetext}
\begin{equation}
	\label{NewPsiVtsd}
	\begin{array}{l}
	{\psi{}}_{vtsd}=\dfrac{V_{\infty{}}}{5}r\left(1-\frac{R^2}{r^2}\right)\left(\cos{\theta{}}-\frac{130}{2Re}\, {\frac {\sin^2{\theta }\cos{\theta} \left( {\frac {
								r}{R}}-1 \right)\frac{{R}^{2}}{{r}^{2}} }{\sqrt {
							\left( 1-{\frac {{R}^{2}}{{r}^{2}}} \right) ^{2}+4\,{\frac {{R}^{2}
									\sin^2{\theta}}{{r}^{2}}}+2}}}\right)\left(\left(\theta{}+\sigma{}\right)\left(\chi\sin{\theta{}}\right)\right)^{\exp{\left[-\frac{V_{\infty}t}{r}\left(1-\frac{R}{r}\right)\right]}}\\[10pt]
				\times {\exp{\left[-\left(\frac{125}{Re}+\frac{r^{2}}{4V_{\infty}^{2}t^{2}}\right)-\left({\exp{(-\frac{130}{Re})}}\left(\frac{r}{\sqrt{R^{2}+4\nu t}}-1\right)^2+\frac{\frac{130}{Re} \sqrt { \left( 1-{\frac {{R}^{2}}{{r}^{2
								}}} \right) ^{2}+4\,{\frac {{R}^{2} \sin^2{\theta} }{{r}^{2}}}+2} \left( {\frac {r(t)}{R}}-1 \right)}{8}\right)\left(\sin^2{\theta{}}+\eta^{\frac{{\exp{\left[-\frac{r^2}{4V_{\infty}^{2}t^{2}}\right]}}}{2}}\right) \right]}}\ln{\left(\dfrac{r}{R}\right)}
				\hspace{3pt},
	\end{array}	
	\end{equation}
\end{widetext}
where
\[
q=\dfrac{130}{Re^2}\left(\dfrac{3.9792Re}{130}\right)^{\exp{\left(-\frac{10}{Re}\right)}}\hspace{3pt},
\]
\[\sigma{}=\left\{\begin{array}{l}-\pi{},\ \ 
\theta{}\geq{}0\ \ \ \  \\
\pi{}, \ \ \theta{}<0\hspace{3pt},\ \ \ \ \ \ \ \ \
\end{array}\right.\],
\[
\chi{}=\left\{\begin{array}{l} 1\ ,\ \ \theta{}\geq{}0\ \
\ \ \ \ \ \ \ \ \ \ \  \\
-1 \ ,\ \ \theta{}<0 \hspace{3pt},\ \ \ \ \ \ \ \ \ \end{array}\right .\]
\[
\eta{}=\left\{\begin{array}{l}1\ \ ,\ \ \theta{}>\frac{\pi{}}{2}\
\vert{}\vert{}\ \ \theta{}<-\frac{\pi{}}{2}\  \\
-1\ ,\ \ \frac{\pi{}}{2}\ \geq{}\ \ \theta{}\geq{}-\frac{\pi{}}{2} \hspace{3pt},\ \ \ \ \ \ \
\end{array}\right. 
\]
\begin{widetext}
\[
r(t)=r\sqrt{\begin{array}{l}$$\left(\sin{\theta(t)}-\frac{ \beta V_{\infty}t}{r}\left(\frac{R^2}{r^2}\sin{\theta(t)}\cos{\theta(t)}\right)\right)^{2}+\left(\cos{\theta(t)}-\frac{ \beta V_{\infty}t}{r}\left(1+\frac{R^2}{r^2}\left(\sin^2{\theta(t)}-\cos^2{\theta(t)}\right)\right)\right)^{2}$$
	\end{array}},
    \]
\end{widetext}
\[
\begin{array}{l}
\theta(t) =\theta-\dfrac{\beta V_{\infty}t\sin{\theta}}{r}\left(1+\dfrac{R^2}{r^2}\right) \hspace{3pt},
\end{array}
\]
and $\beta=-1$.

The coefficient of the exponential expression in Eq.~\ref{NewPsiInv} is the classical zero-circulation stream function that establishes the baseline cylinder flow and describes the streamline of the cylinder surface, $\psi=0$ \cite{Anderson2011}. The exponential argument in this equation (and similar terms in Eqs.~\ref{NewPsiVt}-\ref{NewPsiVtsd}) models the impulsive starting of the motion and approaches one ($1$) as $t$ progresses, fading out its effect \cite[p.~$15$]{Amoloye2024}. 

The established cylinder surface from Eq.~\ref{NewPsiInv} is covered with a surface vortex sheet that is described in Eq.~\ref{NewPsiVt}. This equation has a basic form $\psi_{vt}=({\Gamma}/{2\pi})\ln{(r/R)}$ where $\Gamma$ is the strength or circulation of the vortices. $\Gamma$ is specified by the tangential velocity component of the baseline cylinder flow and is modeled on the Burger's vortex model \cite{Amoloye2020,Amoloye2024}. Burger's vortices are exact, viscous, and unsteady solutions of NSE \cite{VorticityandVortex2006}. Consequently, Eq.~\ref{NewPsiVt} becomes $Re$-dependent \cite[p.~$6$]{Amoloye2024}. 

A parallel sink/source sheet described in Eq.~\ref{NewPsiSS} is placed on the surface vortex sheet so that the introduced singularities are continuously distributed and concentric at every point on the cylinder surface. The basic form of this sink/source sheet is $\psi_{ss}={\Lambda}\theta/2\pi$ where $\Lambda$ is the strength of the sheet. $\Lambda$ is partly determined from the radial velocity component of $\psi_{vt}$ due to the concentricity of the vortices with the sinks/sources \cite[pp. $7$-$15$]{Amoloye2024}.

RPT draws an analogy in the behavior of a fluid particle around the cylinder in the steady closed-wake regime with the two-body problem \cite{Curtis2005} in astrodynamics \cite{Amoloye2024}. This warrants the inclusion of a perifocal stream function, $\psi_p$, in Eq.~\ref{CylGravityVelPot} whose functional form depends on the velocity field of the superposition of Eqs.~\ref{NewPsiInv}-\ref{NewPsiSS}. Equation~\ref{CylGravityVelPot} introduces a wake-analogy factor \cite[pp.~$17$-$19$]{Amoloye2024}, $\mu_g$. 

The vortex shedding stream function, $\psi_{vtsd}$, is modeled on $\psi_{ss}$ which produces the stationary counter-rotating wake bubbles in the flow \cite{Amoloye2024}. The vortices are shed in Eq.~\ref{NewPsiVtsd} with a time-dependent radial location of the wake bubbles, $r(t)$, and a parameter, $\eta$, in the exponential argument that makes Eq.~\ref{NewPsiVtsd} a complex-valued fluctuating function in the rear of the cylinder \cite[pp.~$20$-$25$]{Amoloye2024}. 

$\psi_{vis}$ satisfies the free stream and the wall no-slip boundary conditions with the combined effects of the spatially dominant exponential arguments and $\ln{(r/R)}$. $(r/R -1)$ in the exponential argument of Eq.~\ref{NewPsiVt} together with $\ln{(r/R)}$ ensure that the tangential velocity of Eq.~\ref{NewPsiInv} is counteracted to enforce the no-slip condition on the surface. $\ln{(r/R)}$ also serves to maintain the no-penetration condition on the radial velocity and partially enforces the no-slip on the tangential velocity of the other terms in $\psi_{vis}$ (see Appendix~\ref{Velch} for details). In addition, $\ln{(r/R)}$ ensures that the surface streamline remains $\psi_{vis}=0$, and the negative exponential arguments in $\psi_{vis}$ fade the introduced singularities away from the surface so that the free stream flow is recovered at infinity \cite[p.~$5$]{Amoloye2024}.
\begin{table*}
\begin{centering}
		\caption{Comparative analysis of the properties of $\tilde{\kappa}$ in relation to $\phi$ and $\psi$ \cite{Amoloyeetal2026b}.Reproduced from Physics of Fluids $1$ August $2026$; $38$ ($8$): $086108$, with the permission of AIP Publishing.}
		\label{Function Comparison}
		\begin{tabular}{rrrr}
			\hline
			\hspace{5pt}&$\phi{}$\hspace{20pt} & $\tilde{\kappa}$ &\hspace{20pt}$\psi{}$  \\\hline
			\hspace{5pt}$V_{r}$&$\dfrac{\partial\phi{}}{\partial r}$\hspace{20pt} & $\dfrac{1}{r}\dfrac{\partial \tilde{\kappa{}}}{\partial \theta}$ &\hspace{20pt}$\dfrac{1}{r}\dfrac{\partial \psi{}}{\partial \theta}$  \\
			\\
			\hspace{5pt}$V_{\theta}$&$\dfrac{1}{r}\dfrac{\partial \phi{}}{\partial \theta}$\hspace{20pt} & $-\dfrac{\partial\tilde{\kappa{}}}{\partial r}$ &\hspace{20pt}$-\dfrac{\partial\psi{}}{\partial r}$  \\
			\\
			\hspace{5pt}$V_{z}$&$\dfrac{\partial \phi{}}{\partial z}$\hspace{20pt} & $\dfrac{\partial\tilde{\kappa{}}}{\partial z}$ &\hspace{20pt}$-$  \\
			\\
			\hspace{5pt}$V_x$&$\dfrac{\partial \phi{}}{\partial x}$\hspace{20pt} & $\dfrac{\partial\tilde{\kappa{}}}{\partial x}$ &\hspace{20pt}$\dfrac{\partial\psi{}}{\partial y}$  \\
			\\
			\hspace{5pt}$V_y$&$\dfrac{\partial \phi{}}{\partial y}$\hspace{20pt} & $\dfrac{\partial\tilde{\kappa{}}}{\partial y}$ &\hspace{20pt}$-\dfrac{\partial\psi{}}{\partial x}$  \\
			\\
			\hspace{5pt}$V_{r}$&$V_x\cos{\theta}+V_y\sin{\theta}$\hspace{20pt} & $-V_x\sin{\theta}+V_y\cos{\theta}$ &\hspace{20pt}$V_x\cos{\theta}+V_y\sin{\theta}$  \\
			\\
			\hspace{5pt}$V_{\theta}$&$-V_x\sin{\theta}+V_y\cos{\theta}$\hspace{20pt} & $-\left(V_x\cos{\theta}+V_y\sin{\theta}\right)$ &\hspace{20pt}$-V_x\sin{\theta}+V_y\cos{\theta}$  \\
			\\
			\hspace{5pt}$V_{z}$&$-(V_x\sin{\theta}-V_y\cos{\theta})\cos{\varphi}$\hspace{20pt} & $-\left(V_x\cos{\theta}+V_y\sin{\theta}\right)\cos{\varphi}$ &\hspace{20pt}$-$  \\
			\hline
		\end{tabular}
\end{centering}
\end{table*}

\begin{figure*}
	\centering
	\includegraphics[width=0.5\textwidth,trim={0cm 0cm 0cm 0cm},clip]{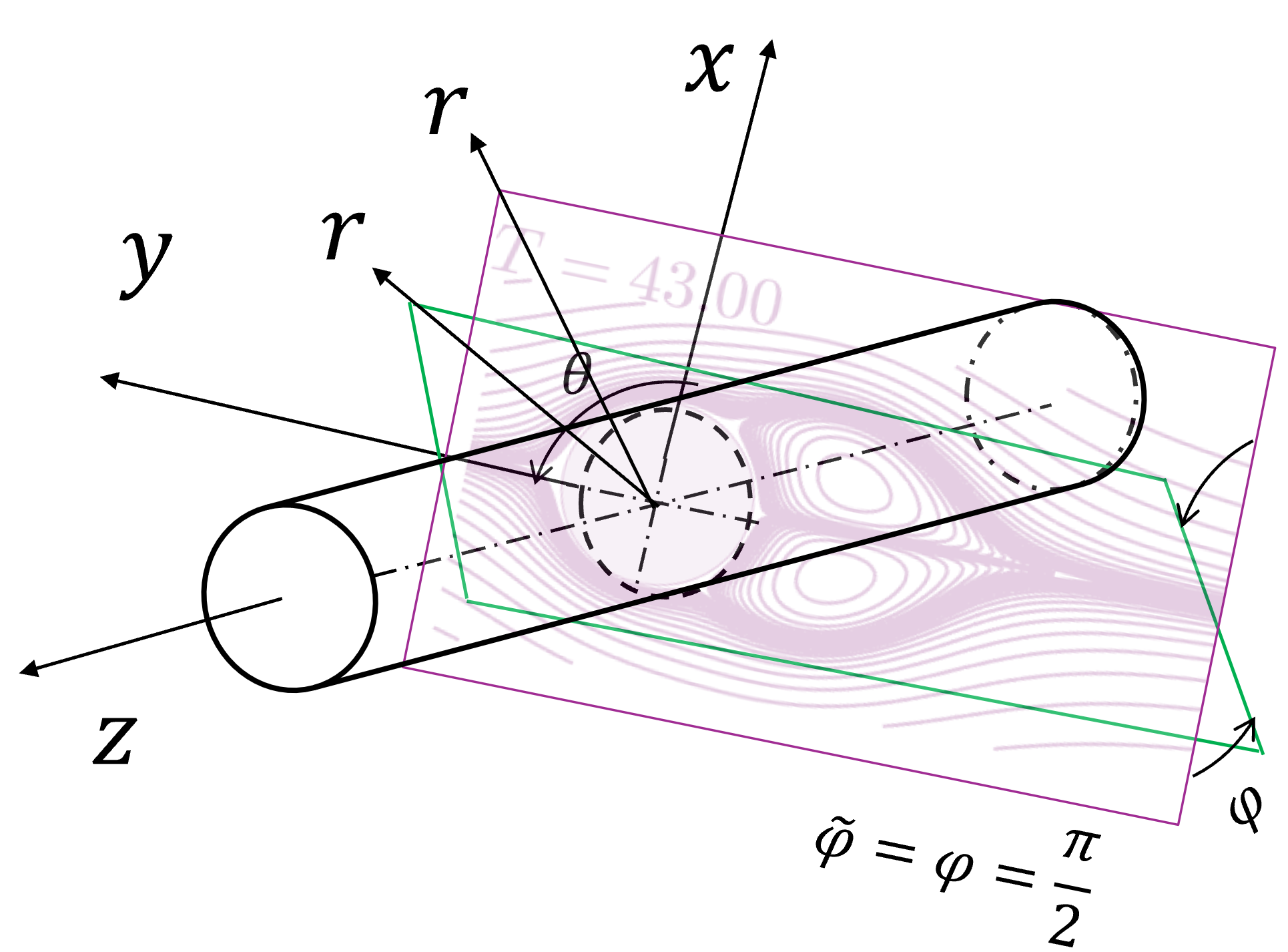}
	\caption{The wind and geometric coordinate axes \color{black}in relation to $\psi_{vis}$ \cite{Amoloye2024}. Reproduced from Physics of Fluids $1$ August $2026$; $38$ ($8$): $086108$, with the permission of AIP Publishing.\color{black}}\label{Fig1}
\end{figure*}

The foregoing description on the $z=0$ plane is two-dimensional, whereas actual cylinder (wake) flows are three-dimensional \cite{Williamson1996}. RPT incorporates three-dimensionality with a distinction between the flow domain axes and the wind axes. Equation~\ref{3Ddomain} describes the fixed Cartesian geometric coordinate axes of the three-dimensional space of the domain, while the Cartesian wind coordinate axes, \color{black}($\tilde{x},\tilde{y},\tilde{z}$), \color{black}are
\begin{equation}
\label{Vwind}
\begin{array}{l}
\tilde{x}=r\cos{\theta}\sin{\tilde{\varphi}} \hspace{10pt}\tilde{y}=r\sin{\theta}\sin{\tilde{\varphi}} \hspace{10pt}\tilde{z}=r\cos{\tilde{\varphi}}=0\\
\tilde{\varphi}=\dfrac{\pi}{2} \hspace{10pt}V_{\tilde{z}}=0 \hspace{10pt}
V_{\tilde{\varphi}}=0\hspace{3pt}.
\end{array}
\end{equation}
\color{black}The radial ($r$) and tangential ($\theta$) variables of the wind axes are the same as those of the domain. $\tilde{\phi}$ and $\tilde{z}$ are azimuthal and spanwise coordinates of the wind axes. \color{black}The wind coordinate system is rotationally symmetric in the three-dimensional space, keeping $\tilde{z}=0$ for all azimuthal planes of the domain. This means that there is a coincident $\tilde{\varphi}=\pi/2$ plane of the wind axes on every $\varphi$ plane of the domain as figure~\ref{Fig1} illustrates. In these planes, $V_{\tilde{z}}=V_{\tilde{\varphi}}=0$. Then, the geometric axes of $\psi_{vis}$ are modified to the wind axes for a full three-dimensional description in the flow domain \cite{Amoloye2024,Amoloyeetal2026b}. 
Using Eq.~\ref{SFVect}, Eq.~\ref{NSEDiff} becomes
\begin{equation}
\label{NSEPsi1a}
\begin{array}{l}
\nabla{}\times{}\dfrac{\partial{}\overrightarrow{\psi{}}}{\partial{}t}+\nabla{}\left(\dfrac{p}{\rho}+\dfrac{|\nabla{}\times\overrightarrow{\psi{}}|^2}{2}-\left[\dfrac{\lambda}{\rho}+2\nu{}\right]\nabla{}\cdot(\nabla{}\times\overrightarrow{\psi{}})\right)\\[10pt]=(\nabla{}\times\overrightarrow{\psi{}})\times{}\omega{}-\color{black}\nu{}\nabla{}\times{}\omega{}\color{black}\hspace{3pt}
\end{array}
\end{equation}
in which there is an obvious difficulty to analytically integrate out the pressure term and conserve momentum. With, $\omega_z=-\nabla^2\psi_{vis}$, there is also a difficulty in satisfying the vorticity transport equation 
\begin{equation}
\label{VortTrnspt}
\begin{array}{l}
\dfrac{\partial{}\omega}{\partial{}t}+(\mathbf{V}\cdot \nabla)\omega = (\omega \cdot \nabla) \mathbf{V} -\omega (\nabla \cdot \mathbf{V}) +  \nu{}\left(\nabla^2 \omega\right)
\end{array}\hspace{3pt} 
\end{equation}
which is a corollary of Eq.~\ref{NSEDiff}.

Equations~\ref{contdiff} and \ref{NSEDiff} are satisfied by a viscous potential flow in which $\omega=0$ \cite{Joseph2006,Joseph2003,Josephetal2007}. Equation~\ref{NSEDiff} emphasizes the impact of $\omega$ within the incompressible Navier-Stokes framework. Although an irrotational velocity field derived from a classical inviscid velocity potential $\mathbf{V} = \nabla \phi$ satisfies Eq.~\ref{NSEDiff} by default, a rotational velocity field generated by a standard planar viscous stream function does not automatically fulfill this condition as Eq.~\ref{NSEPsi1a} shows. To resolve this and satisfy Eqs.~\ref{NSEDiff} and ~\ref{VortTrnspt}, RPT implements the quasi-potential stream function, $\tilde{\kappa}$, in the principal Cartesian coordinate axes of the flow about which $\omega$ is identically equal to zero. This requires clarification because a nonzero $\omega$ will not become zero under any linear transformation. On the principal axes of the strain-rate tensor, there is a local alignment of the coordinate system with the local directions of pure stretching and compression. This zeros out the off-diagonal shear-strain rates, diagonalizing the symmetric tensor with a nonzero trace. Because the vorticity tensor is skew-symmetric, such an alignment cannot eliminate all the off-diagonal terms of the tensor or make the trace of the tensor nonzero. The off-diagonal terms correspond to the components of $\omega$. In addition, a single vector like $\omega$ can physically point in the direction of only one principal strain axis at a time in a three-dimensional space. However, a null vector is an exception to this geometric impossibility. This exception is mathematically achieved in $\tilde{\kappa}$ by transforming $\psi _{vis}$ into the Cartesian wind coordinate system using the following substitution of coordinate variables \cite{Amoloyeetal2026b,Amoloye2024}.

\begin{figure}
	\centering
	\includegraphics[width=0.35\textwidth,trim={0cm 0cm 0cm 0cm},clip]{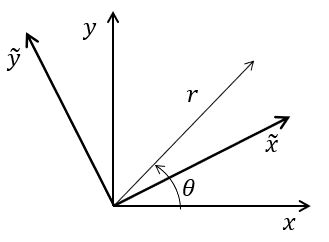}
	\caption{\color{black}The relationship between cylindrical polar and Cartesian coordinate systems in $\psi_{vis}$\cite{Amoloyeetal2026b}. Reproduced from Physics of Fluids $1$ August $2026$; $38$ ($8$): $086108$, with the permission of AIP Publishing.\color{black}}\label{Fig2}
\end{figure}

The standard conversion from Cartesian geometric coordinates $(x, y)$ to cylindrical polar coordinates $(r,\theta)$ is expressed as:
\[r=\sqrt{x^2+y^2}~~~~~~\theta=\arctan{(y,x)}\hspace{3pt}.\]
As shown in figure~\ref{Fig2}, numerous orientations of the wind axes $(\tilde{x}, \tilde{y})$ can map to these identical cylindrical polar coordinates, matching incremental shifts in $\theta $. To establish the principal axes for the flow, the wind axes within $\psi _{vis}$ are rotated so that:
\begin{equation}
\label{kwasutheta}
\theta{}=\arctan{(x,y)}\hspace{3pt}.
\end{equation} 
This alignment ensures that the negative $\tilde{x}$-axis coincides directly with the positive $y$-axis. Through this substitution of variables, $\psi _{vis}$ produces the quasi-potential stream function $\tilde{\kappa}$ along the principal axes \cite{Amoloyeetal2026b,Amoloye2024,Amoloye2020}. In these axes, $\sigma$ is zero,
\[\chi{}=\left\{\begin{array}{l} 1\ ,\ \ \arctan{\left(x,y\right)}\geq{}0\ \
\ \ \ \ \ \ \ \ \ \ \  \\ -1 \ ,\ \ \arctan{\left(x,y\right)}<0\ \ \ \ \ \ \ \ \ \end{array}\right .\hspace{3pt},\] 
\[\eta{}=\left\{\begin{array}{l}1\ \ , \ \ -\frac{\pi{}}{2}\ <\ \ \arctan{\left(x,y\right)}<\frac{\pi{}}{2}\ \\
-1\ ,\ \ \frac{\pi{}}{2}\ \geq{}\ \ \arctan{\left(x,y\right)}\ \ \vert \vert \ \ \arctan{\left(x,y\right)}\leq{}-\frac{\pi{}}{2}\ \ \ \ \ \ \
\end{array}\right. \hspace{3pt},\]
and $\beta$ is positive one $(1)$. Table~\ref{Function Comparison} details the comparative properties of $\tilde{\kappa }$, the classical potential function $\phi$, and a generic stream function $\psi$. In terms of cylindrical polar coordinate variables, $\tilde{\kappa}$ retains an identical structure to $\psi$. However, in the Cartesian coordinate system, the structures are different due to Eq.~\ref{kwasutheta}. This means that the results of differential operations on both $\psi$ and $\tilde{\kappa}$ are agnostic to Eq.~\ref{kwasutheta} and the standard definition of $\theta$ in the cylindrical polar coordinate system, whereas this is not the case in the Cartesian coordinate system.   

\begin{figure}
	\centering
	\subfloat[$\psi=r\sin{\theta}\left(1-\dfrac{R^2}{r^2}\right)$]{\includegraphics[width=0.25\textwidth,trim={0cm 0cm 0cm 0cm},clip]{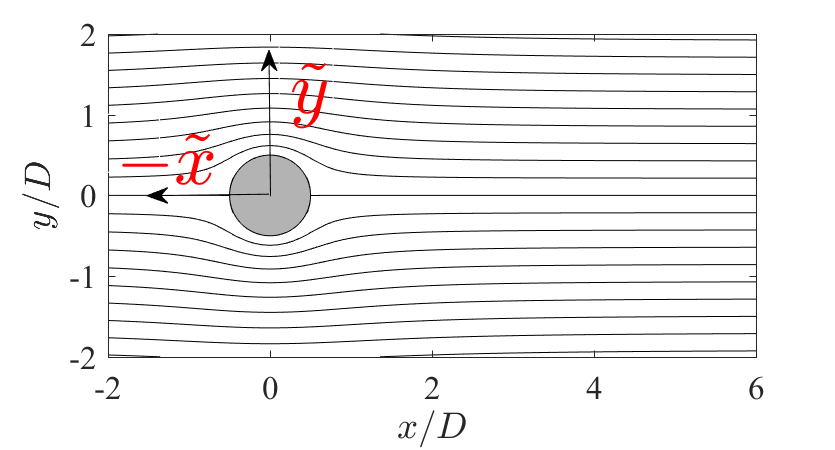}\label{Fig3a}}
	\subfloat[$\phi=r\cos{\theta}\left(1+\dfrac{R^2}{r^2}\right)$]{\includegraphics[width=0.25\textwidth,trim={0cm 0cm 0cm 0cm},clip]{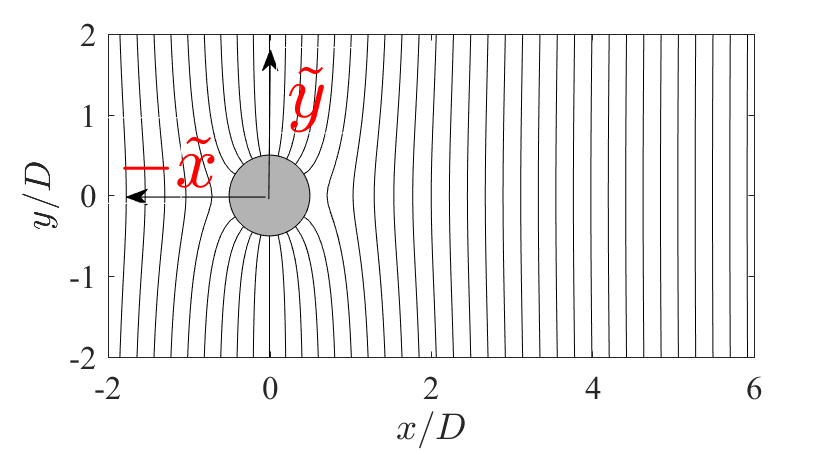}\label{Fig3b}}\\
	\subfloat[$\tilde{\kappa}=r\sin{\left(\arctan{(r\cos{\theta},r\sin{\theta})}\right)}\left(1-\dfrac{R^2}{r^2}\right)$]{\includegraphics[width=0.25\textwidth,trim={0cm 0cm 0cm 0cm},clip]{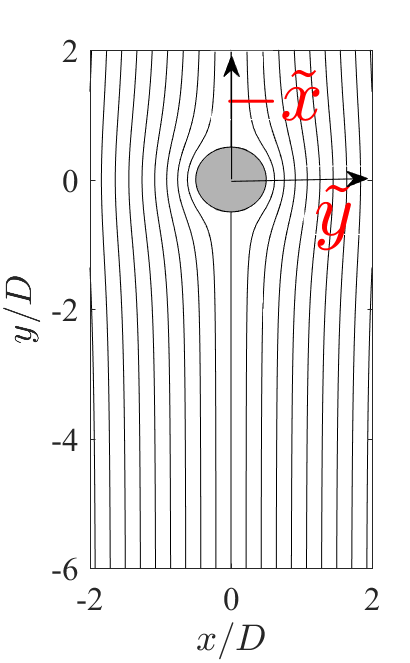}\label{Fig3c}}
	\caption{\color{black}A visual comparison of $\psi$ $(a)$, $\phi$ $(b)$ and $\tilde{\kappa}$ $(c)$ under the inviscid non-lifting cylinder flow conditions \cite{Amoloyeetal2026b}. Reproduced from Physics of Fluids $1$ August $2026$; $38$ ($8$): $086108$, with the permission of AIP Publishing.\color{black}}
	\label{Fig3}
\end{figure}

\begin{figure}
	\centering
	\subfloat[]{\includegraphics[width=0.25\textwidth,trim={0cm 0cm 0cm 0cm},clip]{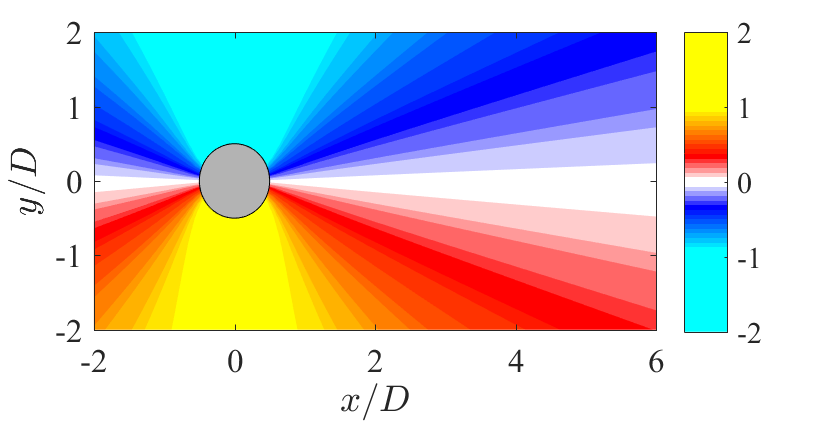}\label{Fig4a}}
	\subfloat[]{\includegraphics[width=0.25\textwidth,trim={0cm 0cm 0cm 0cm},clip]{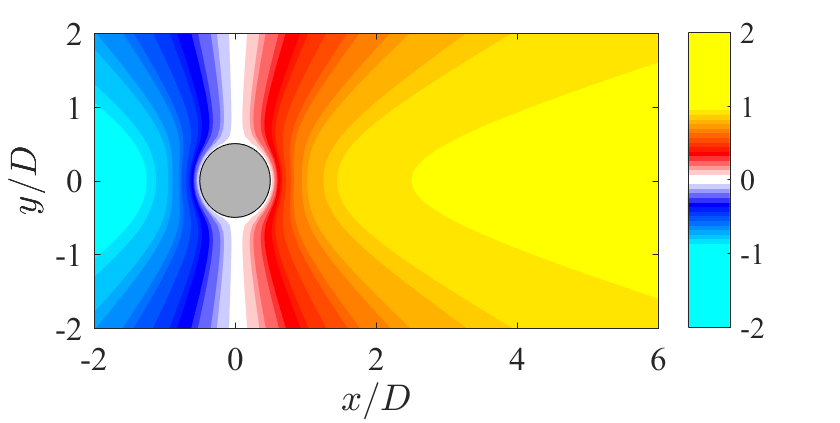}\label{Fig4b}}\\
	\subfloat[]{\includegraphics[width=0.25\textwidth,trim={0cm 0cm 0cm 0cm},clip]{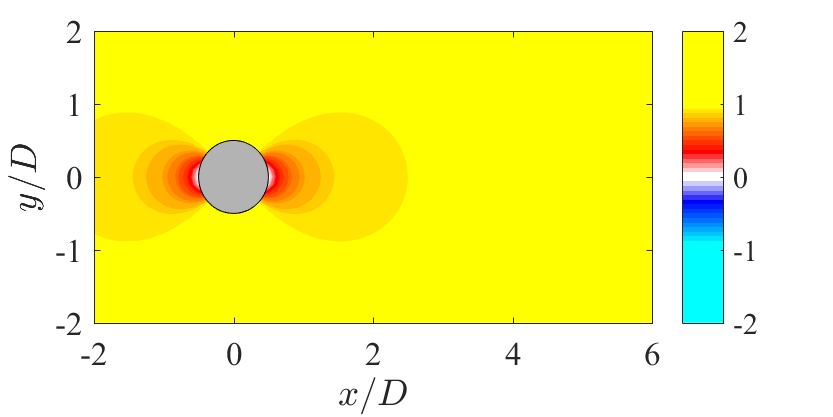}\label{Fig4c}}
	\subfloat[]{\includegraphics[width=0.25\textwidth,trim={0cm 0cm 0cm 0cm},clip]{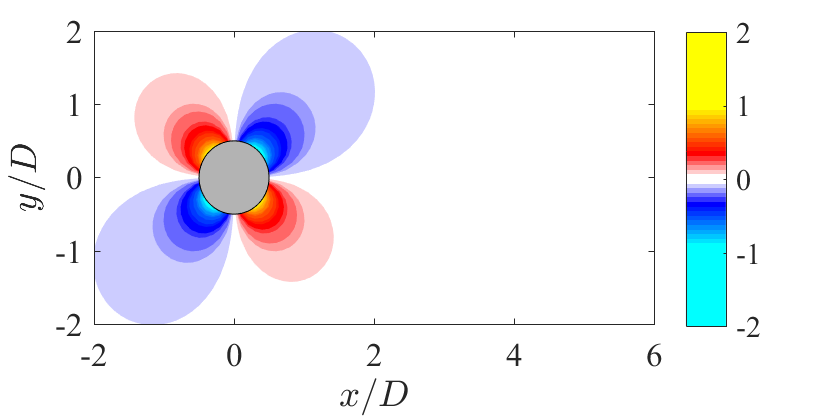}\label{Fig4d}}
	\caption{\color{black}A depiction of the non-dimensional planar velocity field of $\psi$ under the inviscid non-lifting cylinder flow conditions. $(a)$ $V_{\theta}=-{\partial \psi}/{\partial r}$, $(b)$ $V_{r}={\partial \psi}/{r\partial \theta}$, $(c)$ $V_{x}={\partial \psi}/{\partial y}$ and $(d)$ $V_{y}=-{\partial \psi}/{\partial x}$ \cite{Amoloye2021}. Reproduced from Physics of Fluids $1$ August $2026$; $38$ ($8$): $086108$, with the permission of AIP Publishing.\color{black}}
	\label{Fig4}
\end{figure}

\begin{figure}
	\centering
	\subfloat[]{\includegraphics[width=0.25\textwidth,trim={0cm 0cm 0cm 0cm},clip]{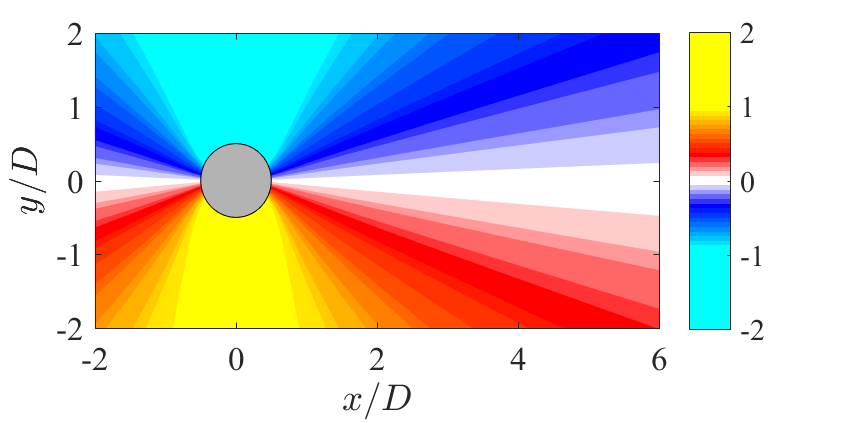}\label{Fig5a}}
	\subfloat[]{\includegraphics[width=0.25\textwidth,trim={0cm 0cm 0cm 0cm},clip]{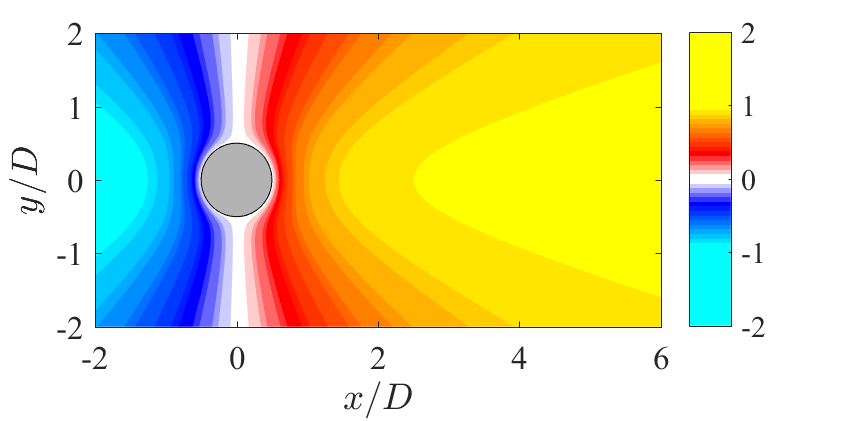}\label{Fig5b}}\\
	\subfloat[]{\includegraphics[width=0.25\textwidth,trim={0cm 0cm 0cm 0cm},clip]{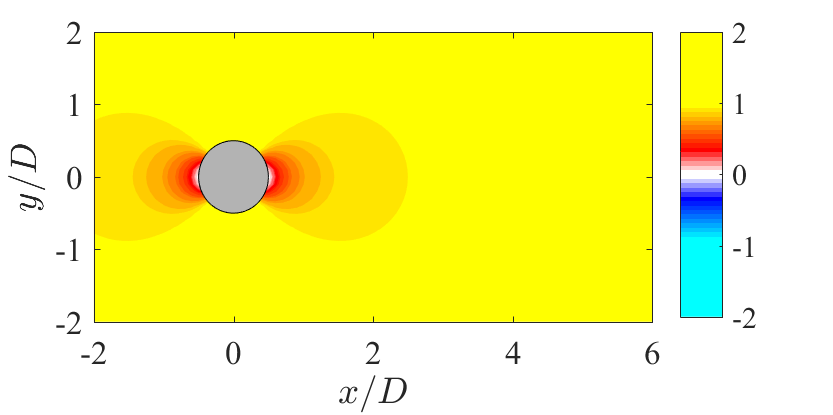}\label{Fig5c}}
	\subfloat[]{\includegraphics[width=0.25\textwidth,trim={0cm 0cm 0cm 0cm},clip]{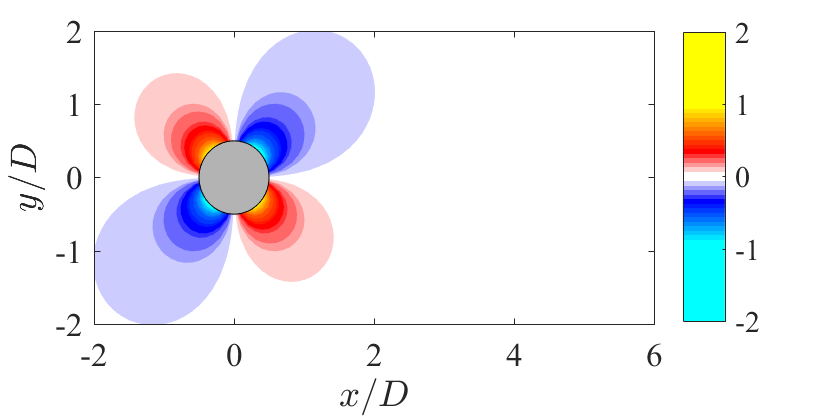}\label{Fig5d}}
	\caption{\color{black}A depiction of the non-dimensional planar velocity field of $\phi$ under the inviscid non-lifting cylinder flow conditions. $(a)$ $V_{\theta}={\partial \phi}/{r\partial \theta}$, $(b)$ $V_{r}={\partial \phi}/{\partial r}$, $(c)$ $V_{x}={\partial \phi}/{\partial x}$ and $(d)$ $V_{y}={\partial \phi}/{\partial y}$ \cite{Amoloye2021}. Reproduced from Physics of Fluids $1$ August $2026$; $38$ ($8$): $086108$, with the permission of AIP Publishing.\color{black}}
	\label{Fig5}
\end{figure}

The idea of a quasi-potential stream function is easily falsifiable under the inviscid, non-lifting cylinder flow conditions. The contours for $\psi$,  $\phi$ and $\tilde{\kappa}$ under these conditions are visualized in figure~\ref{Fig3}.
In the configurations shown in figures~\ref{Fig3a} and~\ref{Fig3b}, the negative $x$-axis aligns with the negative $\tilde{x}$-axis, keeping the freestream flow parallel to both. Conversely, in figure~\ref{Fig3c}, the freestream approaches from the positive $y$-axis. Although the streamlines in figure~\ref{Fig3a} and the potential lines in figure~\ref{Fig3b} are mutually orthogonal, they characterize an identical velocity field. The crucial operational difference lies in differentiation: $\psi$ is differentiated perpendicular to the flow direction to resolve velocity components (as illustrated in figure~\ref{Fig4}), whereas $\phi$ is differentiated parallel to the flow direction (as presented in figure~\ref{Fig5}). Consequently, $\phi$ models an irrotational flow, while $\psi$ can accommodate vorticity \cite{Amoloyeetal2026b,Amoloye2022}.

\begin{figure*}
	\centering
	\subfloat[]{\includegraphics[width=0.25\textwidth,trim={0cm 0cm 0cm 0cm},clip]{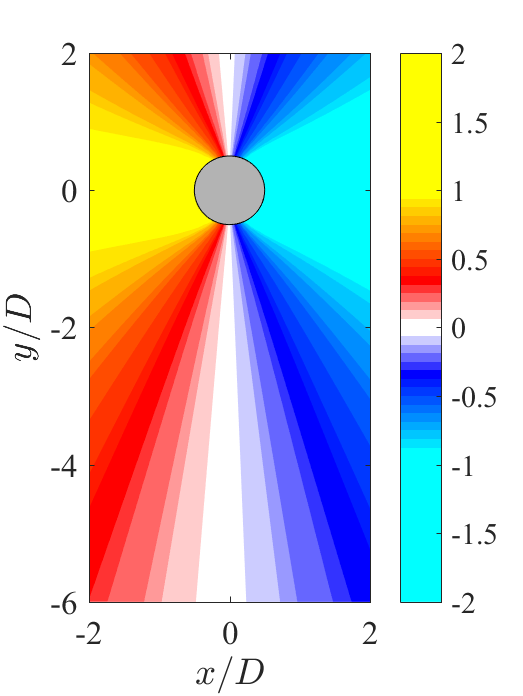}\label{Fig6a}}
	\subfloat[]{\includegraphics[width=0.25\textwidth,trim={0cm 0cm 0cm 0cm},clip]{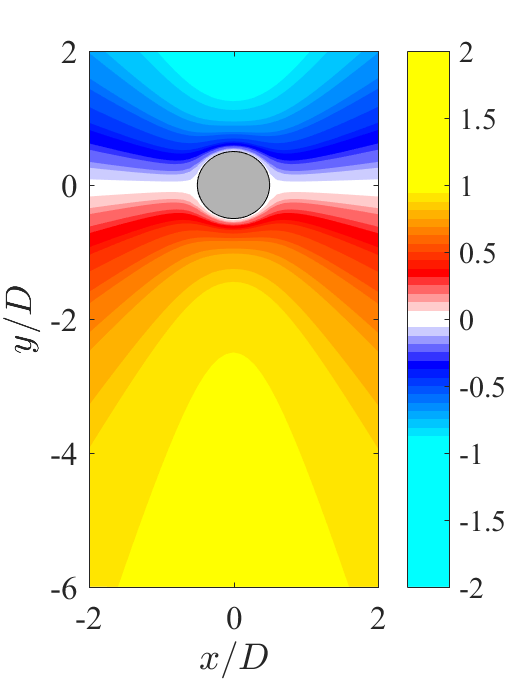}\label{Fig6b}}\\
	\subfloat[]{\includegraphics[width=0.25\textwidth,trim={0cm 0cm 0cm 0cm},clip]{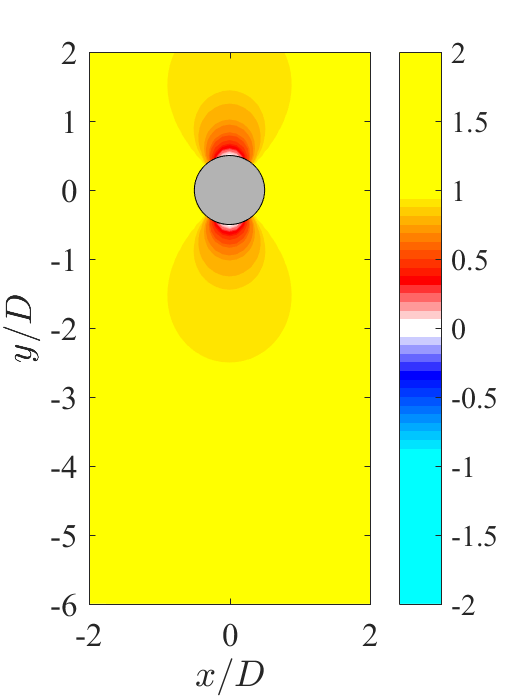}\label{Fig6c}}
	\subfloat[]{\includegraphics[width=0.25\textwidth,trim={0cm 0cm 0cm 0cm},clip]{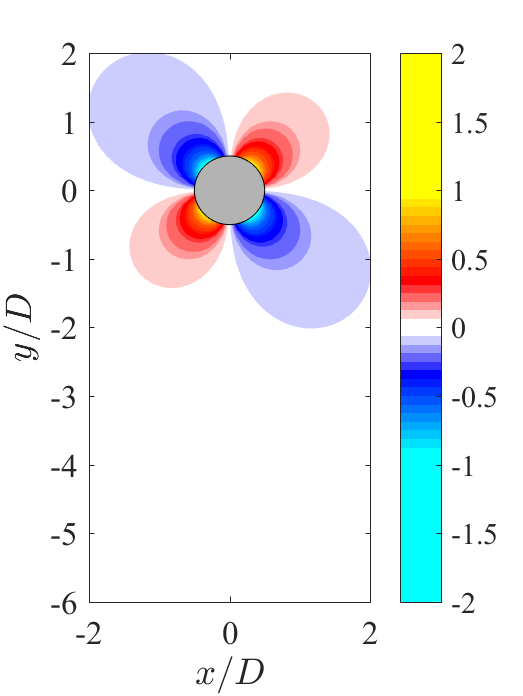}\label{Fig6d}}
	\caption{\color{black}A depiction of the non-dimensional planar velocity field of $\tilde{\kappa}$ under the inviscid non-lifting cylinder flow condition. $(a)$ $V_{\theta}=-{\partial \tilde{\kappa}}/{\partial r}$, $(b)$ $V_{r}={\partial \tilde{\kappa}}/{r\partial \theta}$, $(c)$ $V_{x}={\partial \tilde{\kappa}}/{\partial x}$ and $(d)$ $V_{y}={\partial \tilde{\kappa}}/{\partial y}$ \cite{Amoloye2021}. Reproduced from Physics of Fluids $1$ August $2026$; $38$ ($8$): $086108$, with the permission of AIP Publishing.\color{black}}
	\label{Fig6}
\end{figure*}

\begin{figure*}
	\centering
	\subfloat[]{\includegraphics[width=0.5\textwidth,trim={0cm 0cm 0cm 0cm},clip]{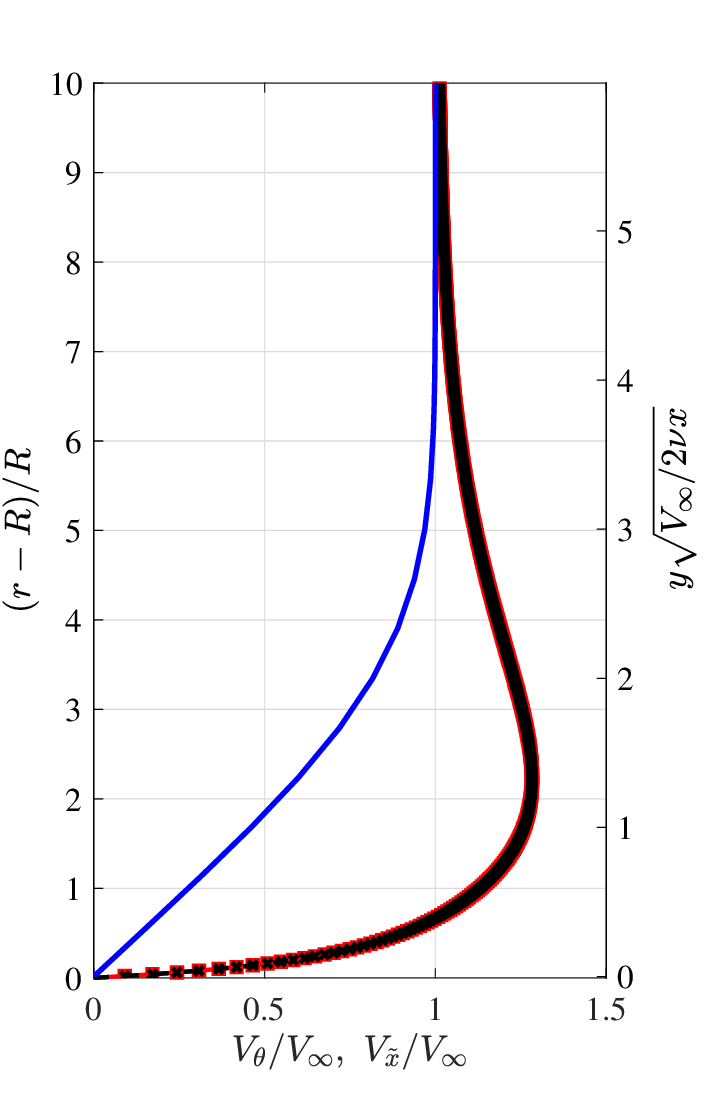}\label{Fig7a}}
	\subfloat[]{\includegraphics[width=0.5\textwidth,trim={0cm 0cm 0cm 0cm},clip]{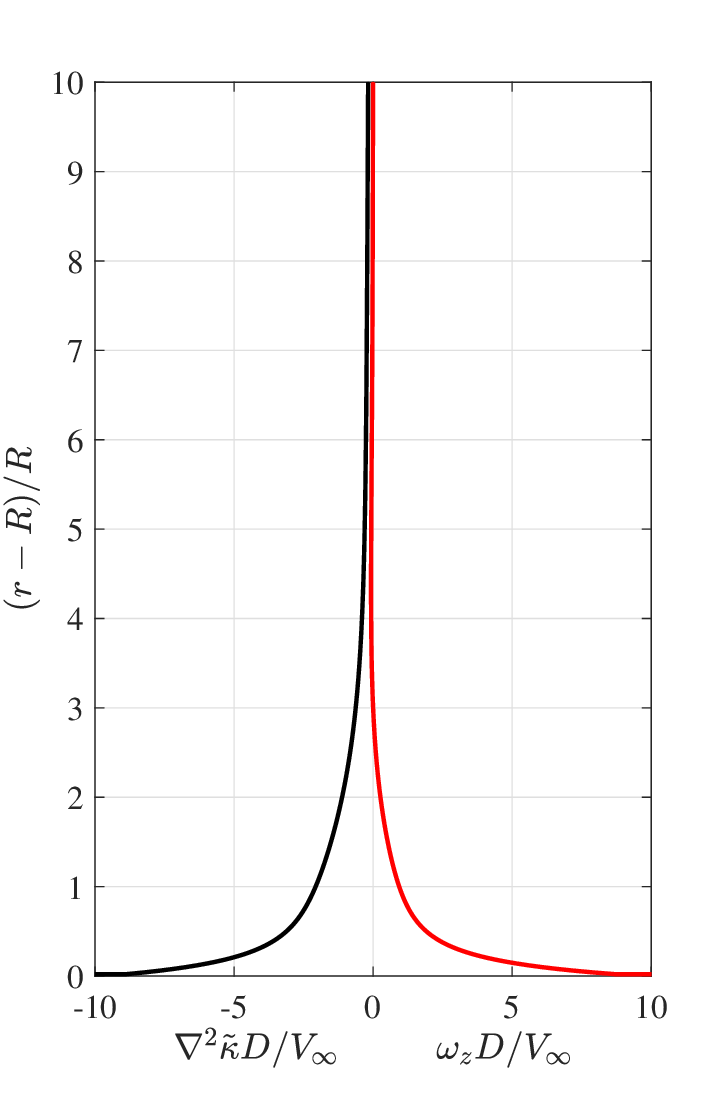}\label{Fig7b}}
	\caption{\color{black} Profiles of selected quantities in a cylinder crest ($\theta=-\pi$) boundary layer for $Re=3,900$, $R=0.047m$, $1.46069\times 10^{-6} m^2s^{-1}$, $\mu_g =6.82\times 10^{-11} m^{3}s^{-2}$ and $T=75,500$. $(a)$ $V_{\theta}/V_{\infty}=-{\partial \tilde{\kappa}}/{\partial r}/V_{\infty}$ (\color{black}$\square-$\color{black}), $V_{\tilde{x}}/V_{\infty}={\partial \tilde{\kappa}}/{\partial x}/V_{\infty}$($--\times$), and Falkner-Skan $V_x/V_{\infty}$ profile for zero -pressure gradient boundary layer (\color{blue}-\color{black}) $(b)$  $\nabla^2\tilde{\kappa}D/V_{\infty}$~$(-)$ and $\omega_zD/V_{\infty}~$(\color{black} $-$\color{black})}
	\label{Fig7}
\end{figure*}

Within the cylindrical polar frame, the velocity components are derived from $\tilde{\kappa}$ in the same manner as a standard stream function $\psi$ (as illustrated in figures~\ref{Fig6a} and~\ref{Fig6b}). However, in the Cartesian wind system, $\tilde{\kappa}$ is defined along the principal axes of a viscous rotational flow where the mathematical expressions for $\omega$ identically equate to zero (see Appendix~\ref{CoMm} for details). These principal axes match the wind axes, directing the negative principal $x$-axis straight into the incoming flow. Operating as a potential function in this frame, the velocity components of $\tilde{\kappa}$ on the $\tilde{z} = 0$ plane in figure~\ref{Fig3c} are evaluated via Eq.~\ref{Vkappa} (as illustrated in figures~\ref{Fig6c} and ~\ref{Fig6d}), where the positive $z$-axis extends outward from the page. 
\begin{equation}
	\label{Vkappa}
	\begin{array}{l}
		x=r\cos{\theta} \hspace{10pt} y=r\sin{\theta} \hspace{10pt}z=r\cos{\varphi}\\[10pt] 
		V_{\tilde{x}}=\dfrac{\partial{}\tilde{\kappa{}}}{\partial{}x}=\dfrac{\partial{}\tilde{\kappa{}}}{\partial{}r}\dfrac{\partial{}r}{\partial{}x}+\dfrac{\partial{}\tilde{\kappa{}}}{r\partial{}\theta{}}\dfrac{r\partial{}\theta{}}{\partial{}x}=-V_{\theta{}}\cos{\theta{}}-V_r\sin{\theta{}}
		\\[10pt]
		V_{\tilde{y}}=\dfrac{\partial{}\tilde{\kappa{}}}{\partial{}y}=\dfrac{\partial{}\tilde{\kappa{}}}{\partial{}r}\dfrac{\partial{}r}{\partial{}y}+\dfrac{\partial{}\tilde{\kappa{}}}{r\partial{}\theta{}}\dfrac{r\partial{}\theta{}}{\partial{}y}=-V_{\theta{}}\sin{\theta{}}+V_r\cos{\theta{}}
		\\[10pt]
		V_{z}=\dfrac{\partial{}\tilde{\kappa{}}}{\partial{}z}=\dfrac{\partial{}\tilde{\kappa{}}}{\partial{}r}\dfrac{\partial{}r}{\partial{}z}=-V_{\theta{}}\cos{\varphi{}}\\
	\end{array}
\end{equation}
Equation~\ref{Vkappa} applies only for a true quasi-potential stream function, taking advantage of Eq.~\ref{kwasutheta}. At the cylinder crests, the streamwise velocity components obtained from $\tilde{\kappa}$ as a stream function in the cylindrical polar coordinate system and as a potential function in the Cartesian coordinate system must be the same. This is the case for the full viscous $\tilde{\kappa}$ as depicted in figure~\ref{Fig7a} in which these velocity components overlap, highlighting its quasi-potential stream function identity. For comparison, the Falkner-Skan boundary layer streamwise velocity profile for zero pressure gradient\cite{White2006} is also plotted on the right vertical axis. These non-dimensional velocity profiles vary from zero ($0$) at the wall to one ($1$) away from the surface, highlighting the viscous boundary conditions. This also shows that the predicted RPT profiles are for the streamwise velocity, confirming the quasi-potential stream function properties of $\tilde{\kappa}$. The basis for comparing the cylinder crest flow with the Falkner-Skan zero-pressure gradient flow is that the wall is parallel to the local freestream for both cases. However, the upstream flow in the cylinder boundary layer just before the crest has a favorable pressure gradient. Consequently, the boundary layer flow at the crest is very accelerated with a steeper velocity profile. The overshoot in the RPT profile is due to curvature effects that make the velocity at the edge of the boundary layer higher than the freestream.         


Despite these quasi-potential properties of $\tilde{\kappa}$, the overall flow field maintains its rotational character, particularly near the surface of the cylinder (see figure~\ref{Fig7b})  and within the wake region \cite{Amoloye2024,Schlichting1979,Amoloye2020}. When strictly analyzed in the cylindrical polar coordinate system, $\tilde{\kappa}$ functions as a stream function, resulting in nonzero $\omega$ components \cite{Amoloye2020,Amoloye2024}. 

Therefore, $\tilde{\kappa}$ satisfies the NSE as \cite{Amoloyeetal2026b}
\begin{equation}
\label{Velpol}
\begin{array}{l}
\mathbf{V}=\nabla{}\tilde{\kappa} (x,y,z)\\[10pt]
\nabla{}\left(\dfrac{\partial{}\tilde{\kappa}}{\partial{}t}+\dfrac{p}{\rho{}}+\dfrac{\left|\nabla \tilde{\kappa}\right|^2}{2}-\left[\dfrac{\lambda}{\rho}+2\nu{}\right]{\nabla{}}^2\tilde{\kappa}\right)=0\hspace{6pt} (t\ge 0).
\hspace{3pt}
\end{array}
\end{equation}
Its irrotational properties as a three-dimensional potential function (in the Cartesian wind coordinate axes) satisfy the inertia and vorticity-dependent viscous terms of the NSE. Having $\omega=0$ in the principal axes is also a solution to the vorticity transport equation (Eq.~\ref{VortTrnspt}). The features of a stream function that it has in the cylindrical polar coordinate wind axes satisfy the vorticity-independent viscous terms of NSE (with the help of the substitution in Eq.~\ref{kwasutheta}). These features also automatically satisfy the continuity equation as  \cite{Amoloye2024,Amoloyeetal2026b}.
\begin{equation}
			\label{contphi}
			\begin{array}{l}
             \overrightarrow{\tilde{\kappa}}= \tilde{\kappa}(r,\theta,\tilde{z})\hat{\mathbf{e}}_{\tilde{z}}\\[10pt]
            \mathbf{V}=\nabla{}\times \overrightarrow{\tilde{\kappa}}\\[10pt]
			\nabla{}\cdot \mathbf{V}=\dfrac{1}{r}\dfrac{\partial (rV_{r})}{\partial r}+\dfrac{1}{r}\dfrac{\partial V_{\theta}}{\partial\theta}+\underbrace{\dfrac{\partial V_{\tilde{z}}}{\partial z}}_{=0}\\[10pt] ~~~~~~~~=0\hspace{10pt} (t\ge 0)
			\end{array}
\end{equation}
to ensure the incompressibilty of the flow at all times. Together, Eqs.~\ref{Velpol} and \ref{contphi} satisfy the Navier-Stokes problem, \color{black}yielding the corresponding pressure field for the conservation of momentum as
\begin{equation}
\label{pressure}
\begin{array}{l}
p=\rho\left[-\left(\dfrac{\partial{}\tilde{\kappa}}{\partial{}t}+\dfrac{V_{\tilde{x}}^2}{2}+\dfrac{V_{\tilde{y}}^2}{2}+\dfrac{V_z^2}{2}\right)\right]\\+\left(\lambda+2\mu\right)\left(\dfrac{{\partial{}}^2\tilde{\kappa}}{\partial{}x^2}+\dfrac{{\partial{}}^2\tilde{\kappa}}{\partial{}y^2}+\dfrac{{\partial{}}^2\tilde{\kappa}}{\partial{}z^2}\right)+ F(t) \hspace{3pt} (t\ge 0).
\end{array}
\end{equation}
$F(t)$ is a function of time. 

Equation~\ref{pressure} should fulfill the hydrostatic condition on the surface of the cylinder where the no-slip and no-penetration conditions hold ($\mathbf{V}=0$). In the hydrostatic condition, the trace of the stress tensor or the sum of normal stresses on the principal axes is tensor invariant \cite{Joseph2003,Joseph2006}. It is defined to be equal to the negative of three times the mechanical pressure, $\tilde{p}$ \cite{White2006}. 
To fulfill this condition, the mean of Eq.~\ref{pressure} is obtained after a rearrangement that is based on factoring out of the deviatoric and isotropic components of the stress term \cite{Amoloyeetal2026b}. This gives
\begin{widetext}
\begin{equation}
	\label{ThermPressure2}
    \begin{array}{l}
	p=-\rho{}\left[\dfrac{\partial{}\tilde{\kappa}}{\partial{}t}+\dfrac{V_{\tilde{x}}^2}{2}+\dfrac{V_{\tilde{y}}^2}{2}+\dfrac{V_z^2}{2}-\dfrac{4}{3}\mu\left(\dfrac{{\partial{}}^2\tilde{\kappa}}{\partial{}x^2}+\dfrac{{\partial{}}^2\tilde{\kappa}}{\partial{}y^2}+\dfrac{{\partial{}}^2\tilde{\kappa}}{\partial{}z^2}\right)\right]+\left(\lambda{}+\dfrac{2}{3}\mu{}\right)\left(\dfrac{{\partial{}}^2\tilde{\kappa}}{\partial{}x^2}+\dfrac{{\partial{}}^2\tilde{\kappa}}{\partial{}y^2}+\dfrac{{\partial{}}^2\tilde{\kappa}}{\partial{}z^2}\right)+F(t)\hspace{3pt}.
    \end{array}
\end{equation}
\end{widetext}
Consequently, the negative of this expression is given as
\begin{widetext}
\begin{equation}
	\label{MechPressure}
	\tilde{p}=\rho \left[\dfrac{\partial \tilde{\kappa}}{\partial{}t}+\dfrac{V_{\tilde{x}}^2}{2}+\dfrac{V_{\tilde{y}}^2}{2}+\dfrac{V_z^2}{2}-\dfrac{4}{3}\mu\left(\dfrac{{\partial{}}^2\tilde{\kappa}}{\partial{}x^2}+\dfrac{{\partial{}}^2\tilde{\kappa}}{\partial{}y^2}+\dfrac{{\partial{}}^2\tilde{\kappa}}{\partial{}z^2}\right)\right]-\left(\lambda{}+\dfrac{2}{3}\mu{}\right)\left(\dfrac{{\partial{}}^2\tilde{\kappa}}{\partial{}x^2}+\dfrac{{\partial{}}^2\tilde{\kappa}}{\partial{}y^2}+\dfrac{{\partial{}}^2\tilde{\kappa}}{\partial{}z^2}\right)-F(t)\hspace{3pt}
\end{equation}
\end{widetext}
where $F(t)$ equals the negative of the total pressure of the free stream, $p_0$.\color{black} 

Owing to the nonzero trace of the strain-rate tensor in the principal axes, the viscous contribution to the pressure field can be determined by evaluating the Laplacian of $\tilde{\kappa}$ in the Cartesian wind coordinate axes, as explicitly derived in Eqs.~\ref{dVxdx} through~\ref{crossderive} in Appendix~\ref{CoMm}. Due to the quasi-potential stream function nature of $\tilde{\kappa}$ acquired via Eq.~\ref{kwasutheta}, this Laplacian is not zero everywhere in the flow field, as figure~\ref{Fig7b} shows. This differentiates Eq.~\ref{pressure} from the unsteady Bernoulli relation\cite{Anderson2011,Schlichting1979} for an irrotational incompressible flow. The ensuing pressure field, Eq.~\ref{MechPressure}, remains invariant under the foregoing substitution of Cartesian coordinate axes variables. Note that figure~\ref{Fig7b} also presents $\omega_zD/V_{\infty}$ which is equal to $-\nabla^2\tilde{\kappa}(r, \theta,\tilde{z})D/V_{\infty}$ in the cylindrical polar wind coordinate axes. The difference between $|\nabla^2\tilde{\kappa}(x,y,z)D/V_{\infty}|$ and $|\nabla^2\tilde{\kappa}(r, \theta,\tilde{z})D/V_{\infty}|$ arises from $\partial V_z/\partial z \ne 0$ and $\partial V_{\tilde{z}}/\partial z=0$. 

\color{black}The generally nonzero Laplacian and vorticity require explicit verification of the quasi-potential simplifications, including the pressure field (Eq.~\ref{pressure}) and residuals. To pass verification, the scalar continuity residual, $R_c$, the vector pressure residual, $\mathbf{R}_{m}$, and the vector curl residual, $\mathbf{R}_{\omega}$, must zero out in the domain, simultaneously. The analysis is presented in detail in Appendix~\ref{ResCh}. All data processing and analyses were performed using custom scripts\cite{taofiqo_2026_24636713} using Maple\cite{Maple} and Matlab\cite{Matlab}. Appendix~\ref{ResCh} shows that $R_c$ is verified. 

$\mathbf{R}_{m}$ is presented in Eq.~\ref{Residual}. Each group of terms in the general arithmetic sum in Eq.~\ref{Residual} is given a numeral subscript. That is, $\mathbf{R}_{m,1}=[\partial \mathbf{V}/\partial t-\nabla(\partial \tilde{\kappa}/\partial t)]$ and so on to the fifth group. $\mathbf{R}_{m,1}$ is consistently below the round-off floor at all points, including the outlier points. $\mathbf{R}_{m,2}$ is not included in the analysis because it is easily assessed to be equal to zero. At all points analyzed, but the outlier points on the center line of the wake, the Lamb vector term, $\mathbf{R}_{m,3}$, equates to zero ($0$) on the surface because of the no-slip and no-penetration conditions. It continues to be zero away from the wall because $\omega$ zeros out. It suffices that $\omega$ equals zero to satisfy $\mathbf{R}_{m,4}$. Verifying that $\omega$ equals zero in the Cartesian coordinate axes also serves to check the curl residual, $\mathbf{R}_{\omega}$. A quantitative description of $\lambda$ continues to be generally difficult \cite{SharmaandKumar2026}. The bulk viscosity, $\left(\lambda{}+2\mu{}/3\right)$, equals to zero ($0$) in Stokes' hypothesis. RPT does not assume the hypothesis but employs a value of $\lambda$ that equates the stagnation pressure coefficient to one ($1$) at the front of the cylinder in the subcritical regime \cite{Amoloye2020}. However, $\mathbf{R}_{m,5}$ is analyzed independently of the viscosity terms with the calculation of $\nabla^{2}\tilde{\kappa}(x,y,z) - \nabla \cdot \mathbf{V}(x,y,z)$.  $\mathbf{R}_{m,5}$ also peaked on the rear axis of the cylinder while remaining below the round-off floor at all other points analyzed. Conclusively, the foregoing quasi-potential simplification is not valid in the outlier points but everywhere else.
\begin{widetext}
\begin{equation}
\label{Residual}
\mathbf{R}_{m} = \left[ \frac{\partial \mathbf{V}}{\partial t} - \nabla \left(\frac{\partial \tilde{\kappa}}{\partial t}\right) \right]+\nabla \left( \dfrac{ V^2}{2} -\dfrac{V_{\tilde{x}}^2}{2}-\dfrac{V_{\tilde{y}}^2}{2}-\dfrac{V_z^2}{2}  \right) - \mathbf{V} \times \omega + \nu (\nabla \times \omega) + \left(\frac{\lambda + 2\mu}{\rho}\right)\left[ \nabla (\nabla^{2}\tilde{\kappa} - \nabla \cdot \mathbf{V}) \right]
\end{equation}
\end{widetext} 
\color{black}
Amoloye \cite{Amoloye2024} provides more details of the theoretical development and the mathematical description of $\tilde{\kappa}$, \color{black}including some constants employed to modulate the flow axis\cite[p.~$33$]{Amoloye2024}. \color{black}That study characterizes RPT for the incompressible flow over an impulsively started circular cylinder ($30 < Re < 10^{4}$, $0.2 \le T \le 77,047$, where non-dimensional time, $T=V_\infty t/R$). The predicted Strouhal number is $0.209$ at $Re = 3,900$, with harmonics and sub-Strouhal frequency fluctuations observed in the temporal energy spectral analysis of the wake flow \cite{Amoloye2024}. Turbulence increases these features, and at $Re = 9,500$, the predictions of the Strouhal number are within $10\%$ of the experimental data \cite{Amoloye2024}. Downstream velocity spectra confirm Kolmogorov’s Five-Thirds law, but limitations remain regarding flow rapidity, vortex core velocity jump, and performance in the laminar regime, which warrant further studies \cite{Amoloye2024}.

Therefore, the effects of the variation of $\nu$ and some other parameters are explored here to address the inadequacies of the theory in the initial transient flow development and flow behavior across wide $Re$ range. 
\section{Results and Discussion}\label{RnD}
Unless otherwise stated, table~\ref{Flow Conditions} presents the conditions that were used for the evaluation. The values for $R$ and $\mu_g$, are the same as those in Amoloye's study \cite{Amoloye2024}. $\nu_\infty$ is arbitrary. All results are for the $z/D=0$ plane, corresponding to a two-dimensional flow. 

\begin{figure}
	\centering
	\includegraphics[width=0.5\textwidth,trim={0cm 0cm 0cm 0cm},clip]{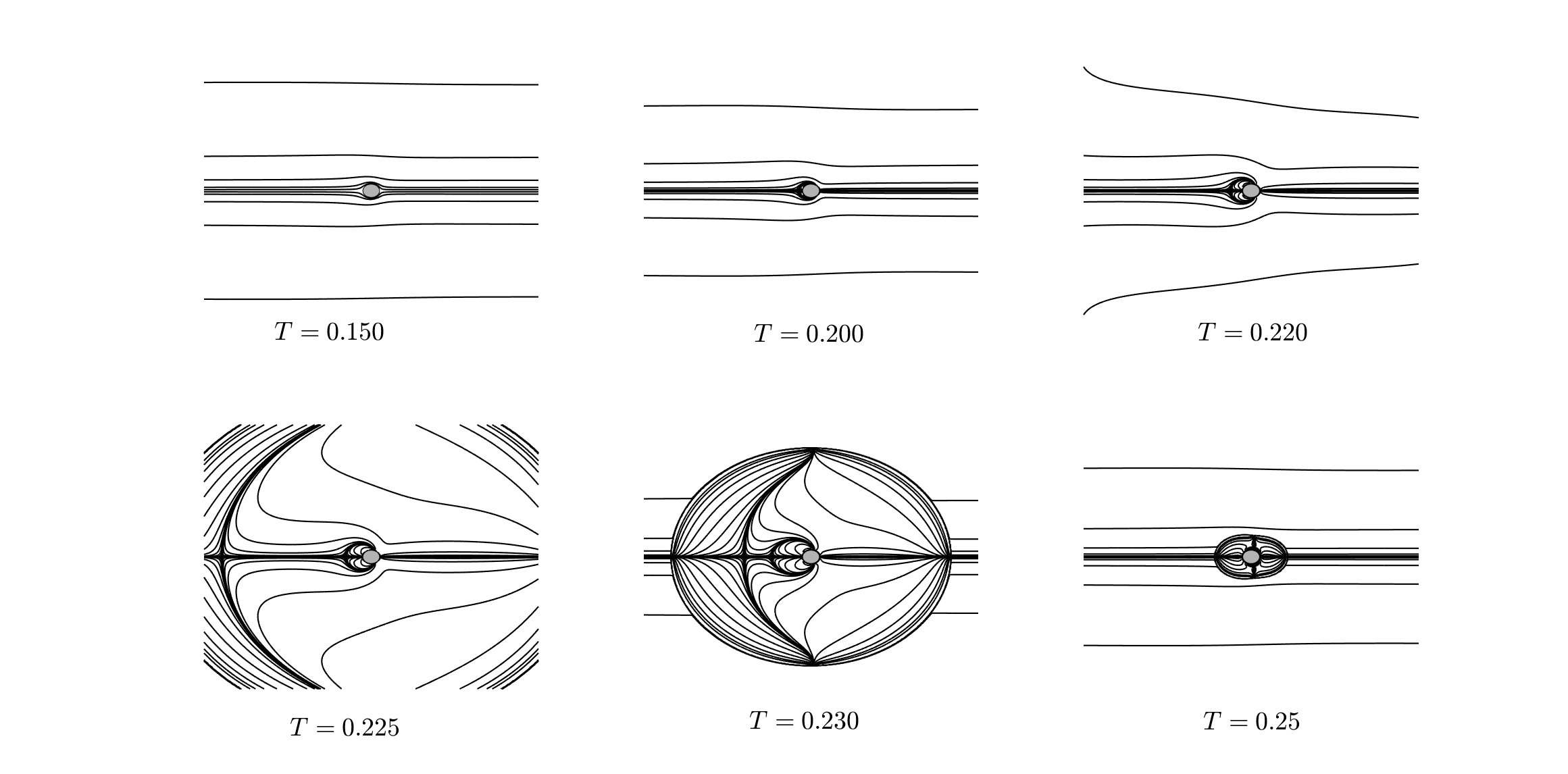}
	\caption{Streamlines of an impulsively started circular cylinder at $Re = 9,500$ and $0.15 \le T \le 0.25$.}\label{Fig8}
\end{figure}

\begin{figure}
	\centering
	\includegraphics[width=0.5\textwidth,trim={0cm 0cm 0cm 0cm},clip]{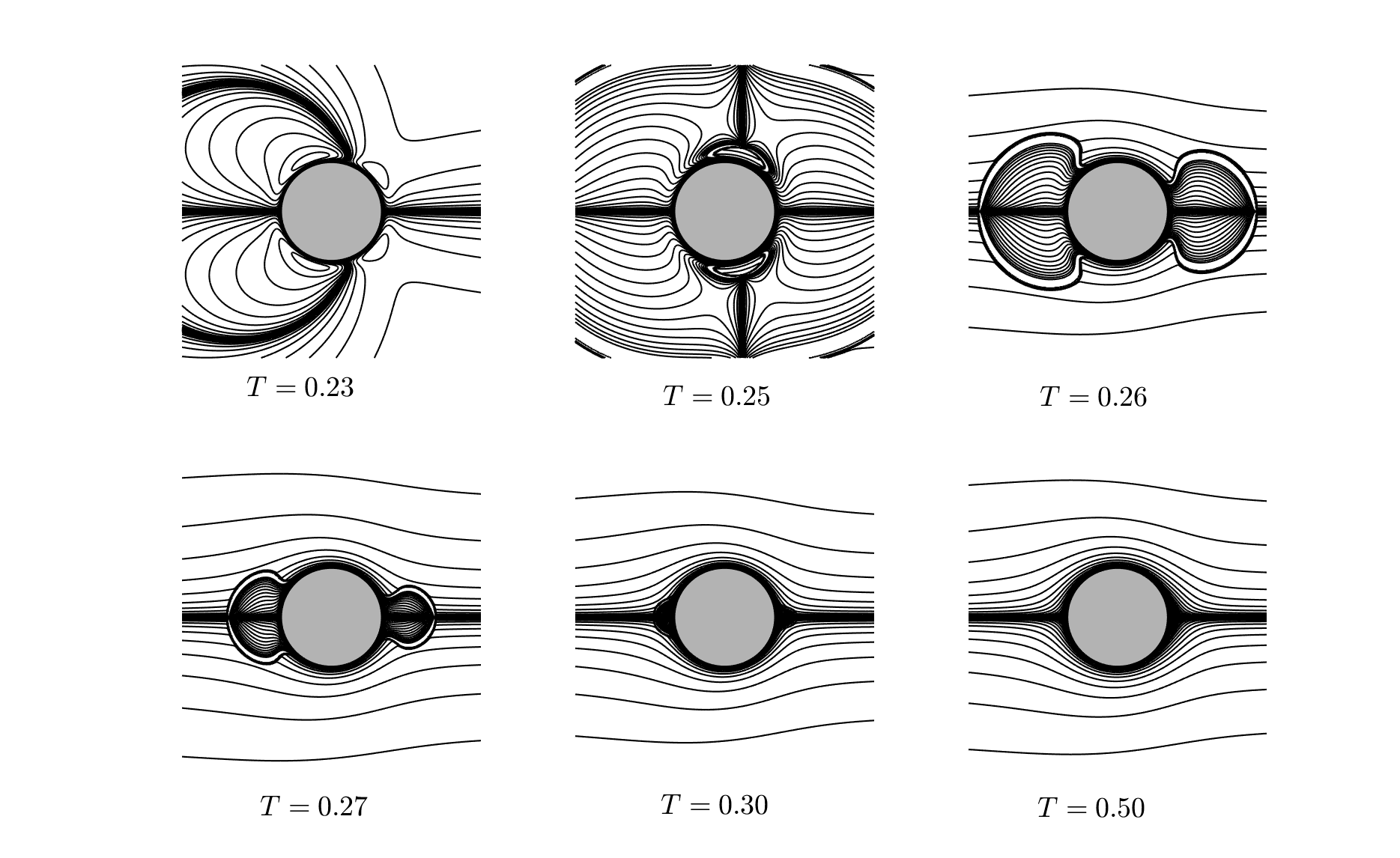}
	\caption{Streamlines of an impulsively started circular cylinder at $Re = 9,500$ and $0.23 \le T \le 0.50$.}\label{Fig9}
\end{figure}

\begin{figure}
	\centering
	\includegraphics[width=0.5\textwidth,trim={0cm 0cm 0cm 0cm},clip]{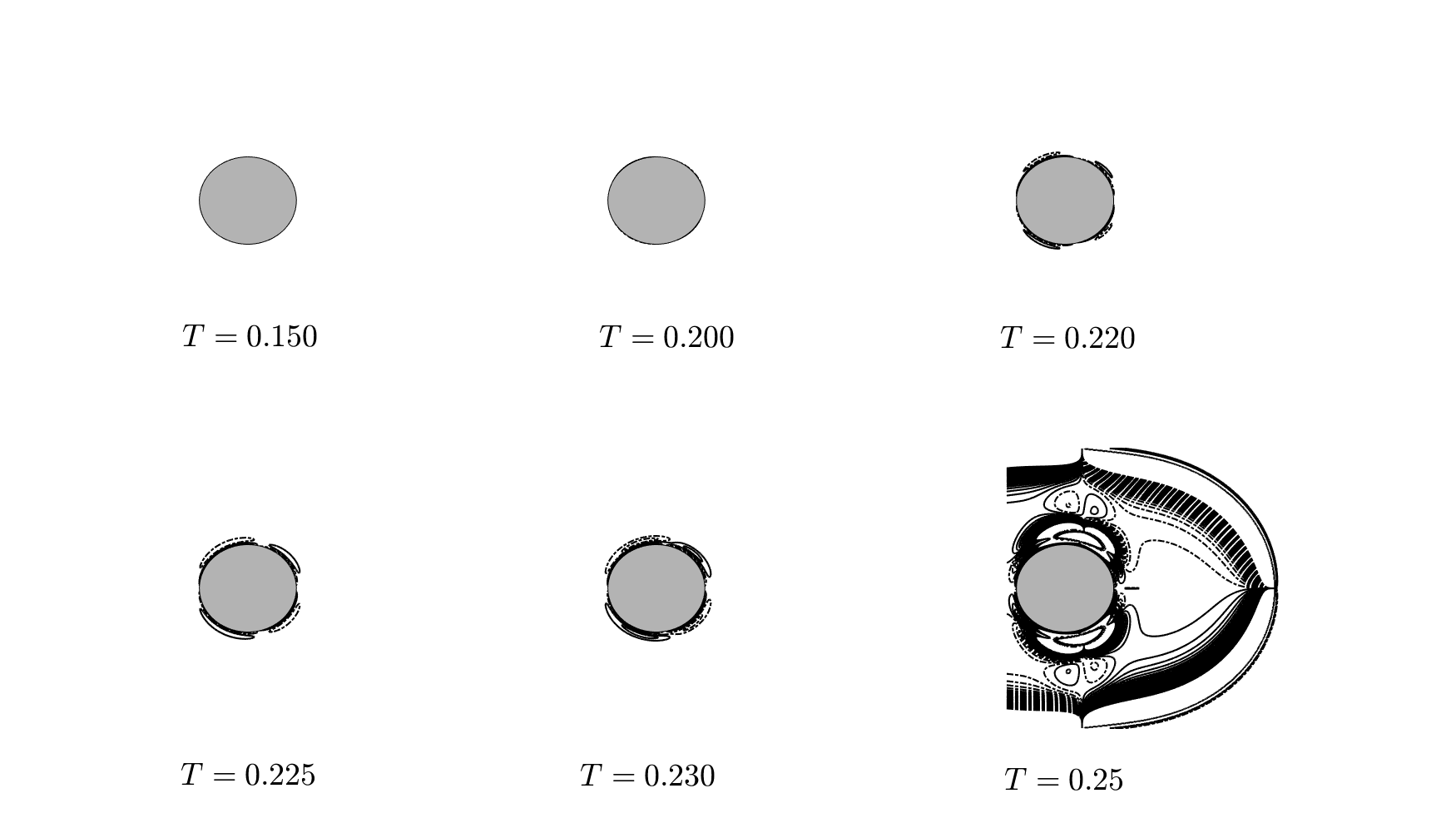}
	\caption{Equi-vorticity contours of an impulsively started circular cylinder at $Re = 9,500$ and $0.15 \le T \le 0.25$.}\label{Fig10}
\end{figure}

\begin{figure}
	\centering
	\includegraphics[width=0.5\textwidth,trim={0cm 0cm 0cm 0cm},clip]{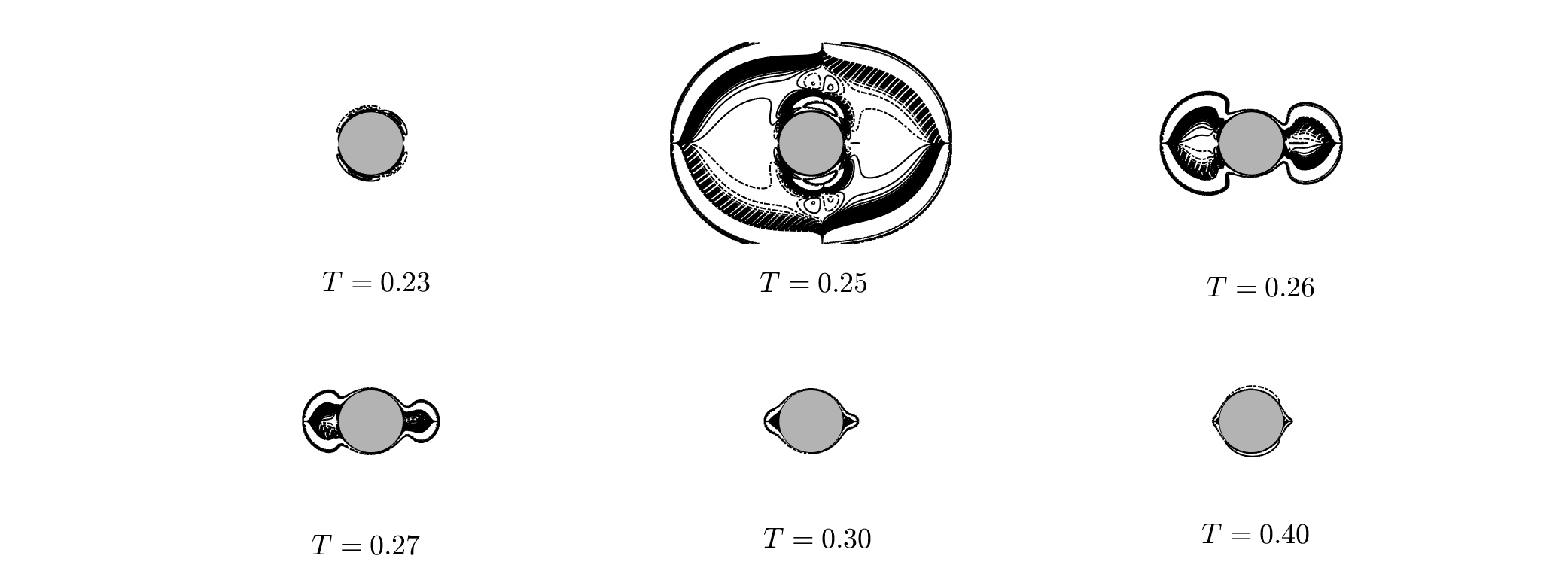}
	\caption{Equi-vorticity contours of an impulsively started circular cylinder at $Re = 9,500$ and $0.23 \le T \le 0.40$.}\label{Fig11}
\end{figure}

 Amoloye suggests that an investigation of RPT results at earlier times than those explored for $Re=9,500$ may reveal the forewake phenomenon \cite{Amoloye2024}. Figures~\ref{Fig8} and \ref{Fig9} present the predicted streamlines at $Re=9,500$ and $0.15 \le T \le 0.5$ for the incompressible flow. Figures~\ref{Fig10} and \ref{Fig11} explore the captured phenomenon with equi-vorticity contours for $0.15 \le T \le 0.4$ and $Re=9,500$. These reveal a multitude of flow features that remind one of acoustic streaming flows \cite{Schlichting1979,VanDyke1982}. A classical potential flow pattern exists at $T=0.15$. Since the flow is starting in a quiescent ambiance, the flow far away from the body at infinity is entirely driven by the oscillation of the boundary layer. The vortex cells in the vicinity of the front of the cylinder are visible at $T=0.2$ with a flow symmetry about the horizontal axis of the cylinder. The vortex cells continue to grow and, by $T=0.22$, the flow far away from the cylinder is visibly distorted. The vortex cells originate from the front and rear stagnating points of the cylinder. Far away from the cylinder, they form a elliptical region separating the region of influence of the oscillating boundary layer from the undisturbed freestream flow that is building up. This elliptical region shrinks towards the cylinder as time progresses to $T=0.25$. A close-up view of the flow in figure~\ref{Fig9} reveals the eyes of the vortices around the cylinder at $T=0.23$. By $T=0.25$, a pocket of a steady streaming flow emanating from the cylinder crests dislodges the vortex eyes radially outward from the cylinder. At $T=0.26$, the constituent parts of the shrinking elliptical region dissociate, with each portion receding towards the respective stagnation point by $T=0.5$.      
 
\begin{table}
	\begin{center}
		\caption{Cylinder radius and flow conditions.}
		\label{Flow Conditions}
		\begin{tabular}{rrrr}
			\hline
			 Radius, $R$ ($m$) &  $\nu_{\infty}$ ($m^{2}s^{-1}$) & $\mu_g$ ($m^{3}s^{-2}$) &   \\\hline
			$0.047$&  $1.46069\times 10^{-8}$ &$6.82\times 10^{-11}$ &\\
			\hline
		\end{tabular}
	\end{center}
\end{table}

\begin{figure}
	\centering
	\includegraphics[width=0.5\textwidth,trim={0cm 0cm 0cm 0cm},clip]{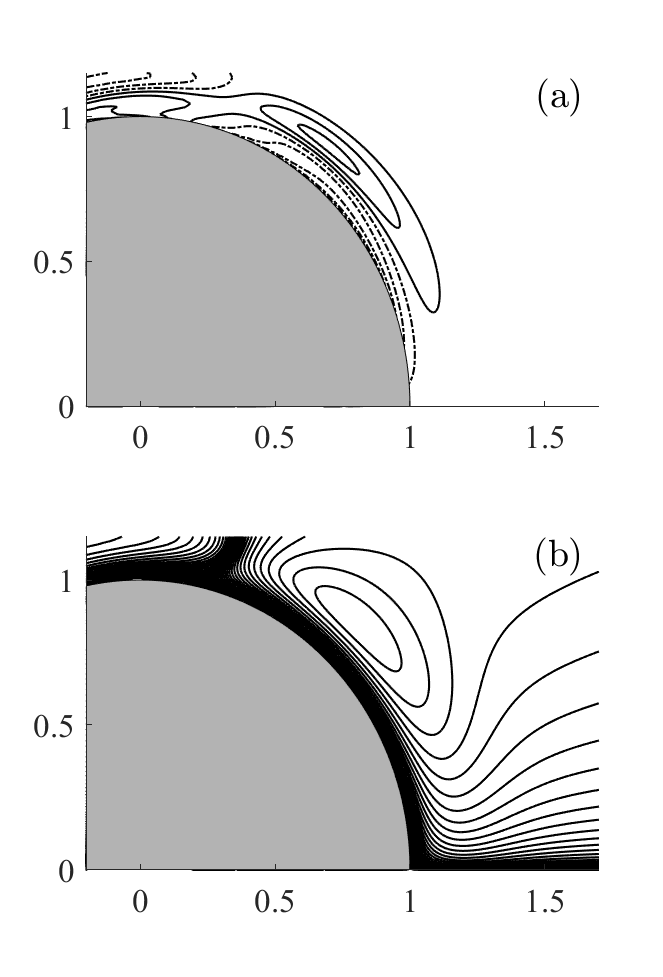}
	\caption{RPT (a) equi-vorticity lines and (b)instantaneous streamlines at $Re = 9,500$ and $T=0.2356$.}\label{Fig12}
\end{figure}

Although the preceding discussion is interesting, it is important to determine whether the captured phenomenon is an adequate representation of the experimentally observed forewake. Figure~\ref{Fig12} shows the equi-vorticity of RPT and the instantaneous streamline contours of the incompressible flow at $Re=9,500$ and $T = 0.2356$. These are compared with CFD and experimental observation of the forewake also at $Re=9,500$ but later at $T=2.0$ \cite[figure 25, p.~28]{KoumoutsakosandLeonard1995}. The rapid vortex \cite{BouardandCout1980} is visible and similar in the equi-vorticity contours, but the difference of its mechanism is obvious from the dissimilarity in the streamline contours. The similarity between the streamlines in figure~\ref{Fig9} (at $T=0.23$) and the streamlines of the secondary flow around an oscillating cylinder \cite[figure 2, p.~455]{Nurievetal2018} suggests that the phenomenon captured in the RPT analysis is an artifact of the exponential expressions used to model the impulsive starting of the flow \cite[pp.~$15$-$17$]{Amoloye2024}. However, the introduced low-amplitude oscillations are ephemeral similarly to the forewake \cite{BouardandCout1980}. By $T=0.5$, the forewake has completely disappeared, and the main wake has started forming.  

\begin{figure}
	\centering
	\includegraphics[width=0.5\textwidth,trim={0cm 0cm 0cm 0cm},clip]{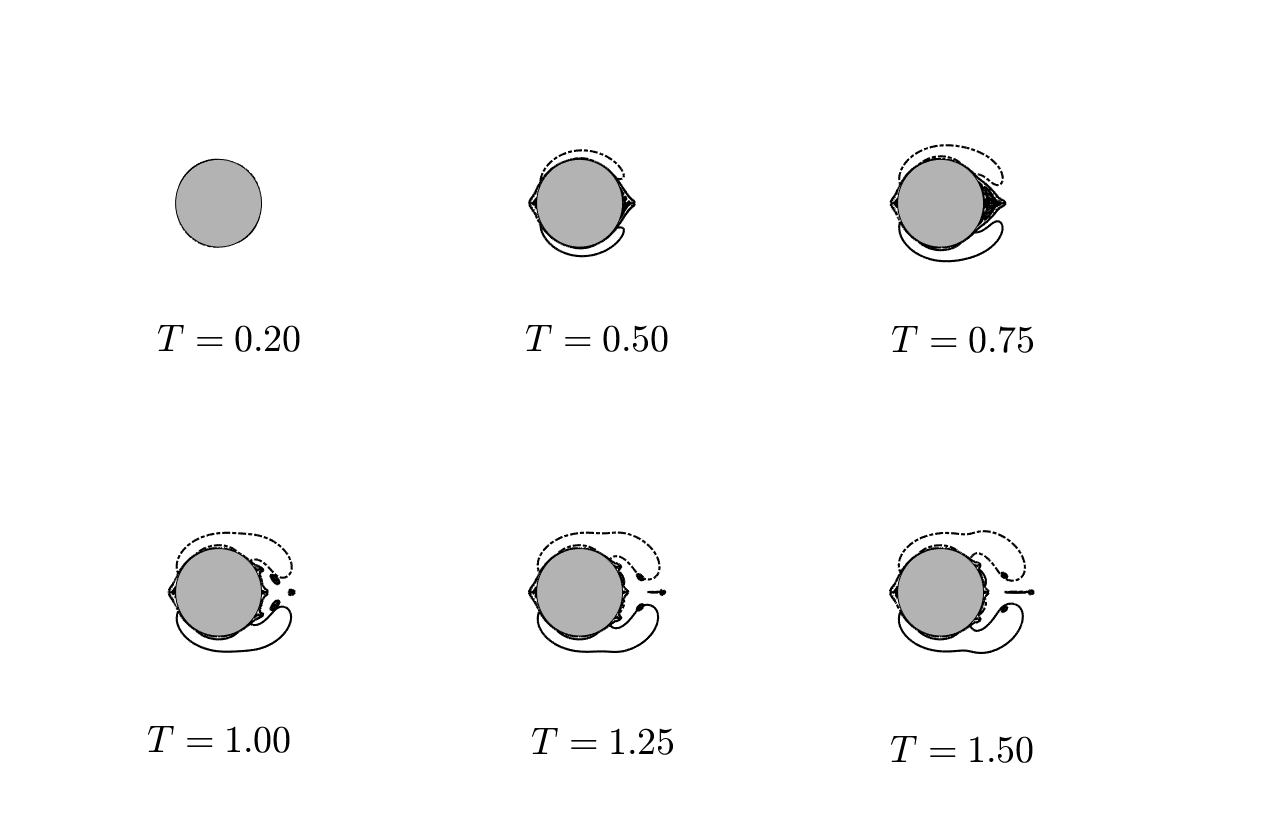}
	\caption{Equi-vorticity contours of an impulsively started circular cylinder at $Re = 9,500$ and $0.20 \le T \le 1.50$.}\label{Fig13}
\end{figure}

\begin{figure}
	\centering
	\includegraphics[width=0.5\textwidth,trim={0cm 0cm 0cm 0cm},clip]{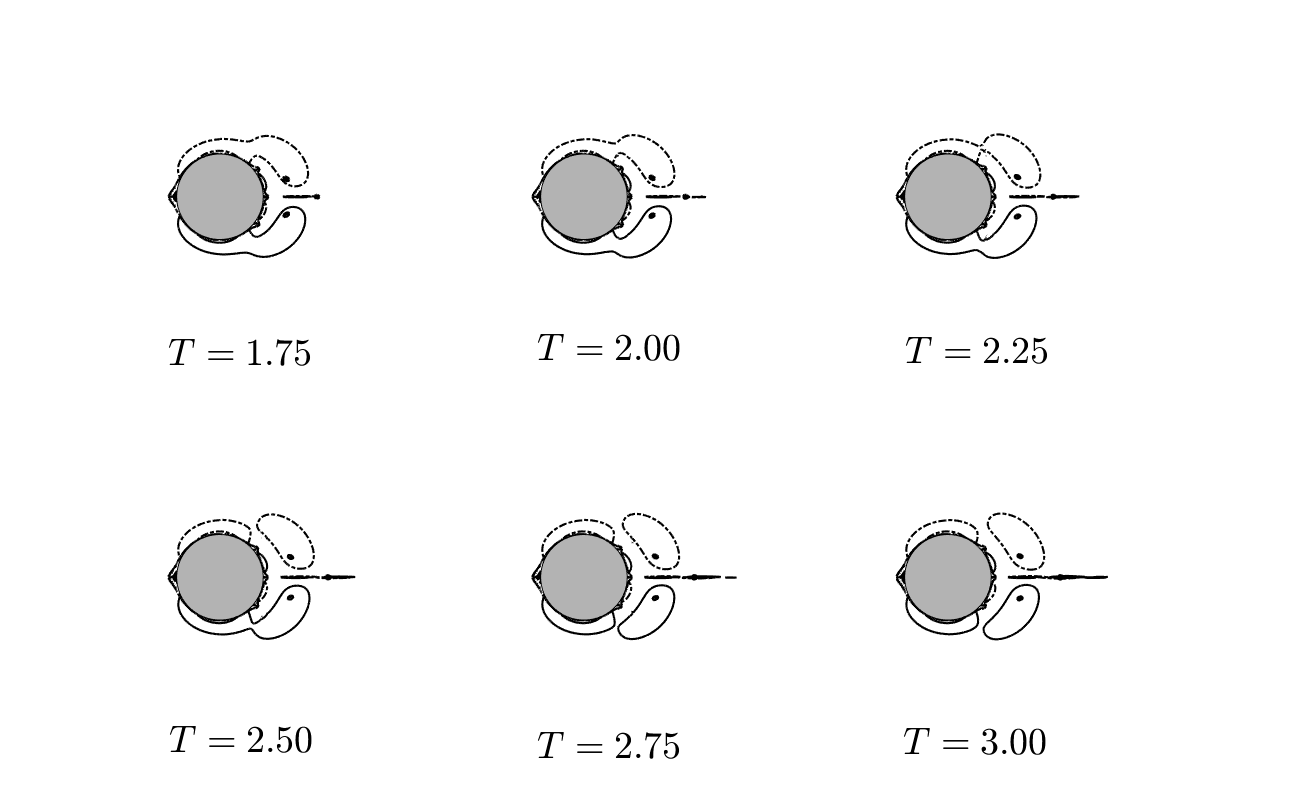}
	\caption{Equi-vorticity contours of an impulsively started circular cylinder at $Re = 9,500$ and $1.75 \le T \le 3.00$.}\label{Fig14}
\end{figure}

\begin{figure}
	\centering
	\includegraphics[width=0.5\textwidth,trim={0cm 0cm 0cm 0cm},clip]{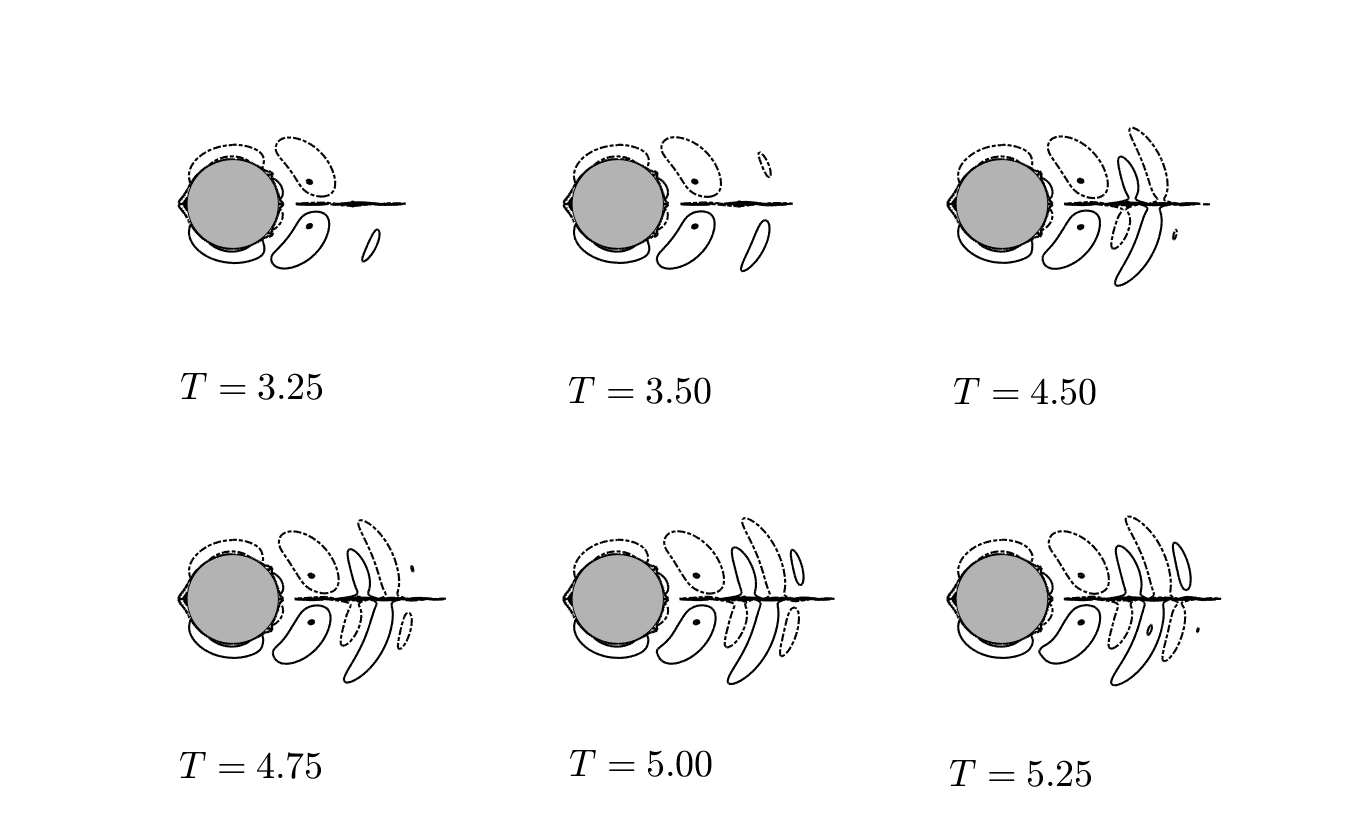}
	\caption{Equi-vorticity contours of an impulsively started circular cylinder at $Re = 9,500$ and $3.25 \le T \le 5.25$.}\label{Fig15}
\end{figure}

\begin{figure}
	\centering
	\includegraphics[width=0.5\textwidth,trim={0cm 0cm 0cm 0cm},clip]{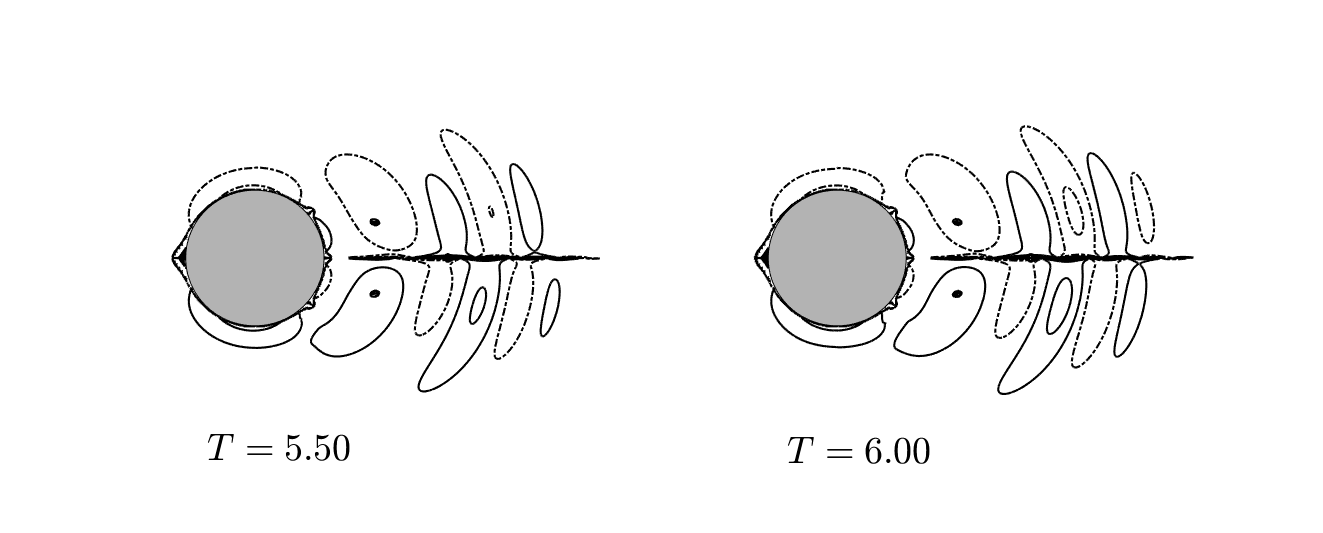}
	\caption{Equi-vorticity contours of an impulsively started circular cylinder at $Re = 9,500$ and $5.00 \le T \le 6.00$.}\label{Fig16}
\end{figure}

Figures~\ref{Fig13} to \ref{Fig16} show the equi-vorticity contours of the incompressible flow at $Re=9,500$ and $0.20 \le T \le 6.00$. Compared with Amoloye's analysis \cite[figures~$50$-$53$, pp.~$49$-$50$]{Amoloye2024}, the present analysis stabilizes with a value of $\nu$ that is three orders of magnitude smaller. Stable flow symmetry over time is seen to be promoted by decreasing the value of $\nu$ while keeping $R$ constant. A similar effect is obtained by increasing $R$ by orders of magnitude while keeping $\nu$ constant. Conversely, decreasing the value of $R$ by orders of magnitude from Amoloye's anlaysis \cite{Amoloye2024} further destabilizes the flow and introduces smaller scale structures. For a specific $Re$, there are infinite possibilities of flow patterns depending on the values of $\nu$ and $R$ because these variables also independently feature in the \textit{Kwasu} function ($\tilde{\kappa}$) \cite{Amoloye2024,Amoloye2018,Amoloye2020} alongside $Re$. Varying these variables to match the flow pattern observed experimentally is synonymous with adjustment of artificial viscosity in CFD simulations \cite{MargolinandLloyd2023}. The consequent values of $V_\infty$ and $T$ depend on these variables. This is why closer agreement can be observed in some $Re$ than in others while maintaining the same values of $\nu$ and $R$ in a wide range of flow regimes \cite{Amoloye2024}. Additionally, it is the reason for the mismatch in the non-dimensional timing in the initial transient flow development.

\begin{figure}
	\centering
	\includegraphics[width=0.5\textwidth,trim={0cm 0cm 0cm 0cm},clip]{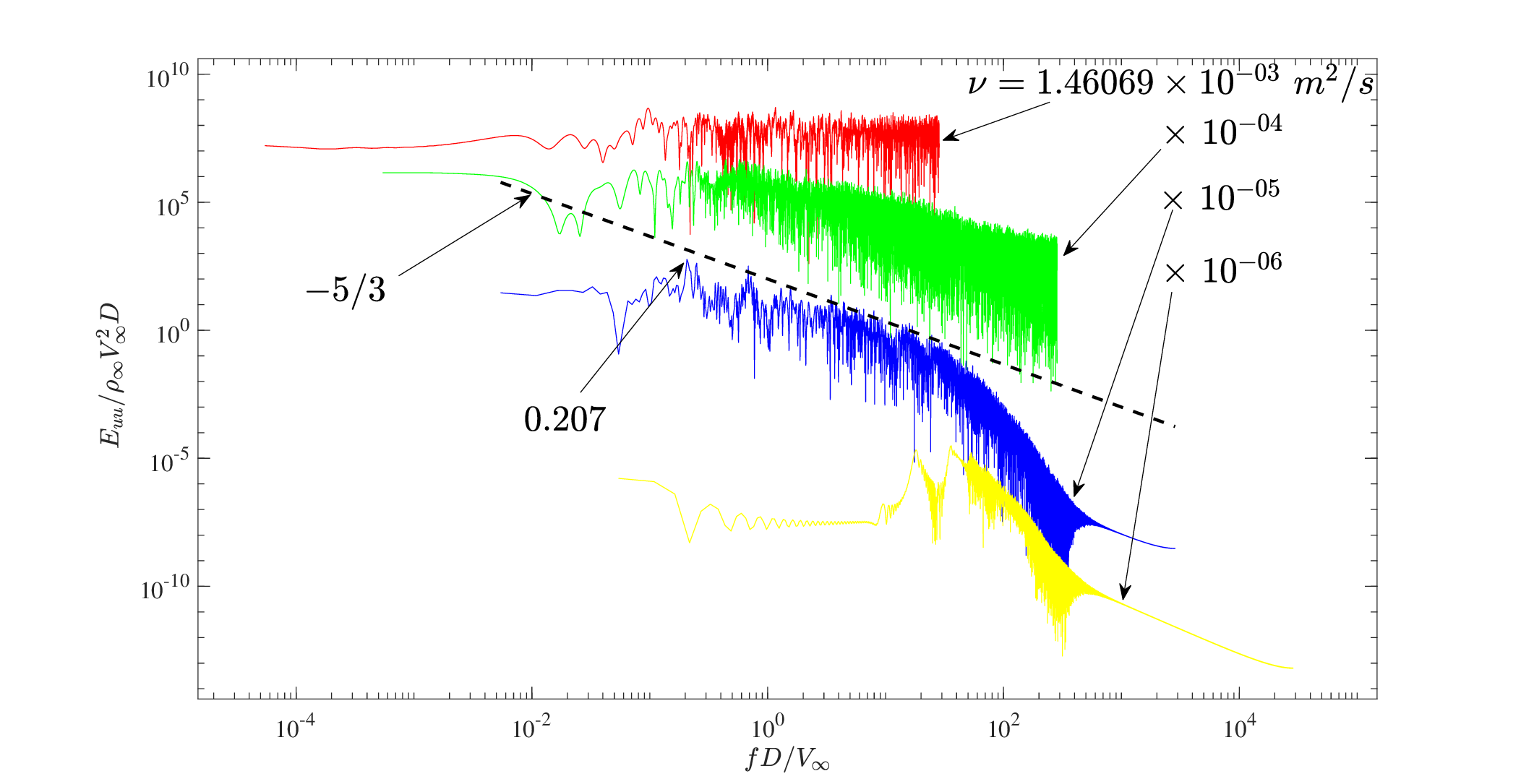}
	\caption{Energy spectra of the streamwise velocity in a cylinder wake at $Re = 3,900$ and $(\tilde{x}/D,\tilde{y}/D)=(0.69,0.69)$.}\label{Fig17}
\end{figure}

Figure~\ref{Fig17} presents one-dimensional energy spectra of the streamwise velocity in a cylinder wake for four values of viscosity at $Re=3,900$ and $(\tilde{x}/D,\tilde{y}/D)=(0.69,0.69)$. These are wavenumber (spatial) spectra obtained in the same way as Amoloye's spectra analysis \cite{Amoloye2024,Amoloyeetal2026b}, but in terms of the Strouhal frequency $fD/V_\infty$ (where $f$ is the vortex shedding frequency). From $\nu=1.46069\times 10^{-3} m^2/s$, the spectra are appropriately displaced downward from the previous one. So, figure~\ref{Fig17} does not necessarily illustrate any apparent changes in the magnitudes of the spectra with viscosity. At $\nu=1.46069\times 10^{-5} m^2/s$, the predicted Strouhal number is $0.207$. This agrees excellently with $0.209$ obtained from RPT temporal spectra analysis and experimental and CFD data \cite{Amoloye2024}. The turbulence decay rate also obeys Kolmogorov's Five-Thirds law. Gradually increasing the viscosity by orders of magnitude increasingly redistributes the energy of the spectra. Lower frequency oscillations are activated at the expense of the smallest scale features, moving the spectra to the left. The spectra also become shallower because the higher frequencies are energized. The broad-banded spectra peaks indicate that vortex shedding still occurs but at sub-Strouhal frequencies. Reducing the viscosity by an order of magnitude has opposite effects to these. Very small scale and high-frequency structures are activated in place of large scale features, extending the spectra to the right. However, the highest frequency oscillations are de-energized. This makes the spectra steeper. There is a valley around the Strouhal frequency, and all the spectra peaks occur at very high frequencies. This suggests that the flow has no apparent vortex shedding and is dominated by shear layer instabilities.      

\section{Conclusion}\label{C}
This paper discusses RPT applied to the incompressible flow over an impulsively started circular cylinder, a classical fluid dynamics problem with significant engineering applications. 

RPT builds on previous theoretical, numerical, and experimental research on the cylinder flow to refine the classical potential flow theory. It incorporates viscous effects, flow fluctuations, and three-dimensional influences in its formulation. The continuity and Navier-Stokes equations are used to model the flow, with boundary conditions at the free stream and cylinder surface. RPT uses a quasi-potential viscous stream function to satisfy the Navier-Stokes problem\color{black}, analysis of the residuals shows that the quasi-potential simplification is valid for points in the domain excluding outlier points in the wake. \color{black}A previous study on RPT explores flow features at different Reynolds numbers and non-dimensional times. Identifying phenomena such as vortex shedding and turbulence decay, the model aligns with experimental and CFD results. It predicts wake characteristics, including Strouhal numbers and turbulence features. However, discrepancies remain in flow rapidity, vortex core velocity, and laminar regime performance.

The present analysis shows that varying the viscosity and radius of the cylinder significantly affects the stability, symmetry, and energy spectra of the wake flow. The lower viscosity promotes smaller-scale structures and higher-frequency oscillations, while the higher viscosity redistributes energy to lower frequencies and larger-scale features. This highlights the need for careful parameter selection in evaluations. The study suggests that RPT can be improved by addressing the identified viscosity-dependent issues, potentially through adjustments similar to artificial viscosity in CFD simulations.

Overall, the research contributes to the understanding of RPT and its application to modeling complex fluid dynamics phenomena, while identifying areas for further investigation and improvement.

\begin{acknowledgments}
This work is based on Ph.D. research supervised by Professor Marilyn J. Smith and funded by Kwara State University, Malete, Nigeria. The premise of the research was first presented at $2018$ AIAA SciTech Forum (AIAA 2018-1288). 
\end{acknowledgments}
\section*{Data Availability Statement}
Data supporting the findings of this study are available in the article.
\section*{Conflict of Interest}
The authors declare that they have no conflict of interest.

\bibliography{references}
\appendix
\section{Cancellation Check on the Wall Velocity}\label{Velch}
Equation~\ref{PsiInvVel} is the velocity field of $\psi_{inv}$. It retains both the impermeability and the slip conditions on the inviscid flow for the velocity at the wall where $r=R$.   
\begin{equation}
	\label{PsiInvVel}
	\begin{array}{l}
    {V_{r_{inv}}}= V_{\infty{}}\cos{\theta{}}\left(1-\frac{R^2}{r^2}\right)\exp{\left[-\frac{R^2}{4V^2_\infty t^2} \right]}\\
	V_{\theta_{inv}} = -V_{\infty{}}\sin{\theta{}}\left(1+\frac{R^2}{r^2}\right)\exp{\left[-\frac{R^2}{4V^2_\infty t^2} \right]}
				\hspace{3pt}.
	\end{array}	
	\end{equation}
Equation~\ref{PsiVtVel} presents the velocity components of $\psi_{vt}$ with particular focus on $\ln{(r/R)}$, whose value is equal to zero ($0$) at the wall. Since $\ln{(r/R)}$ remains constant with a differentiation in the tangential direction to obtain the radial velocity, it ensures the impermeability condition. 
\begin{widetext}
\begin{equation}
	\label{PsiVtVel}
	\begin{array}{l}
    {V}_{r_{vt}}=-\dfrac{\partial}{\partial \theta}\left(V_{\infty}\sin{\theta}\left(1+\frac{R^2}{r^2}\right) {\exp{\left[-\left(\frac{r^{2}}{4V_{\infty}^{2}t^{2}}+\frac{qRe\sqrt { \left( 1-{\frac {{R}^{2}}{{r}^{2}}} \right) ^{2}+4\,{\frac {{R}^{2} \sin^2{\theta} }{{r}^{2}}}+2}\left( {\frac {r}{R}}-1 \right)}{8} \right) \right]}}\ln{\left(\dfrac{r}{R}\right)}\right)\\
	{V}_{\theta_{vt}}=V_{\infty}\sin{\theta}\left(1+\frac{R^2}{r^2}\right) {\exp{\left[-\left(\frac{r^{2}}{4V_{\infty}^{2}t^{2}}+\frac{qRe\sqrt { \left( 1-{\frac {{R}^{2}}{{r}^{2}}} \right) ^{2}+4\,{\frac {{R}^{2} \sin^2{\theta} }{{r}^{2}}}+2}\left( {\frac {r}{R}}-1 \right)}{8} \right) \right]}}\\[10pt]
    +\dfrac{\partial}{\partial r}\left(V_{\infty}r\sin{\theta}\left(1+\frac{R^2}{r^2}\right) {\exp{\left[-\left(\frac{r^{2}}{4V_{\infty}^{2}t^{2}}+\frac{qRe\sqrt { \left( 1-{\frac {{R}^{2}}{{r}^{2}}} \right) ^{2}+4\,{\frac {{R}^{2} \sin^2{\theta} }{{r}^{2}}}+2}\left( {\frac {r}{R}}-1 \right)}{8} \right) \right]}}\right)\ln{\left(\dfrac{r}{R}\right)}
				\hspace{3pt},
	\end{array}	
	\end{equation}
\end{widetext}
Using the product rule to differentiate $\ln{(r/R)}$ and its coefficient terms in $\psi_{vt}$ to obtain $V_{\theta_{vt}}$, there are two blocks of terms. One retains $\ln{(r/R)}$ and goes to zero on the wall. The other block goes to zero by counteracting $V_{\theta_{inv}}$ as
\begin{widetext}
\begin{equation}
	\label{PsiInvVtVel}
	\begin{array}{l}
    V_{\theta_{inv}}+V_{\theta_{vt}}=-V_{\infty{}}\sin{\theta{}}\left(1+\frac{R^2}{R^2}\right)\exp{\left[-\frac{R^2}{4V^2_\infty t^2} \right]}+V_{\infty}\sin{\theta}\left(1+\frac{R^2}{R^2}\right) {\exp{\left[-\left(\frac{R^{2}}{4V_{\infty}^{2}t^{2}}+\frac{qRe\sqrt { \left( 1-{\frac {{R}^{2}}{{R}^{2}}} \right) ^{2}+4\,{\frac {{R}^{2} \sin^2{\theta} }{{R}^{2}}}+2}\left( {\frac {R}{R}}-1 \right)}{8} \right) \right]}}\\[10pt]
	~~~~~~~~~~~~~~~~~= -2V_{\infty{}}\sin{\theta{}}\exp{\left[-\frac{R^2}{4V^2_\infty t^2} \right]}+2V_{\infty}\sin{\theta}{\exp{\left[-\frac{R^{2}}{4V_{\infty}^{2}t^{2}} \right]}}\\[10pt]
    ~~~~~~~~~~~~~~~~~=0
				\hspace{3pt}.
	\end{array}	
	\end{equation}
\end{widetext} 

The radial and tangential velocity components of $\psi_{ss}$ are
\begin{widetext}
\begin{equation}
	\label{PsiSSVel}
	\begin{array}{l}
	{V}_{r_{ss}}=\dfrac{\partial}{\partial \theta}\left(V_{\infty{}}\left(1-\frac{R^2}{r^2}\right)\left(\theta{}+\sigma{}\right)\left(\chi\sin{\theta{}}\right)^{\exp{\left[-\frac{\sqrt{\nu t}}{r}\left(1-\frac{R}{r}\right)\right]}}\right.\\[10pt]\times\left.\left(\cos{\theta{}}-\frac{qRe}{2}\, {\frac {\sin^2{\theta }\cos{\theta} \left( {\frac {r}{R}}-1 \right)\frac{{R}^{2}}{{r}^{2}} }{\sqrt {\left( 1-{\frac {{R}^{2}}{{r}^{2}}} \right) ^{2}+4\,{\frac {{R}^{2}\sin^2{\theta}}{{r}^{2}}}+2}}}\right) {\exp{\left[-\left(\frac{r^{2}}{4V_{\infty}^{2}t^{2}}+\frac{qRe\sqrt { \left( 1-{\frac {{R}^{2}}{{r}^{2}}} \right) ^{2}+4\,{\frac {{R}^{2} \sin^2{\theta} }{{r}^{2}}}+2} \left( {\frac {r}{R}}-1 \right)}{8}\right) \right]}}\ln{\left(\dfrac{r}{R}\right)}\right)\\[10pt]
    {V}_{\theta_{ss}}= - V_{\infty{}}\left(1-\frac{R^2}{r^2}\right)\left(\theta{}+\sigma{}\right)\left(\chi\sin{\theta{}}\right)^{\exp{\left[-\frac{\sqrt{\nu t}}{r}\left(1-\frac{R}{r}\right)\right]}}\\[10pt]\times\left(\cos{\theta{}}-\frac{qRe}{2}\, {\frac {\sin^2{\theta }\cos{\theta} \left( {\frac {r}{R}}-1 \right)\frac{{R}^{2}}{{r}^{2}} }{\sqrt {\left( 1-{\frac {{R}^{2}}{{r}^{2}}} \right) ^{2}+4\,{\frac {{R}^{2} \sin^2{\theta}}{{r}^{2}}}+2}}}\right) {\exp{\left[-\left(\frac{r^{2}}{4V_{\infty}^{2}t^{2}}+\frac{qRe\sqrt { \left( 1-{\frac {{R}^{2}}{{r}^{2}}} \right) ^{2}+4\,{\frac {{R}^{2} \sin^2{\theta} }{{r}^{2}}}+2} \left( {\frac {r}{R}}-1 \right)}{8}\right) \right]}}\\[10pt]
    -\dfrac{\partial}{\partial r}\left(V_{\infty{}}r\left(1-\frac{R^2}{r^2}\right)\left(\theta{}+\sigma{}\right)\left(\chi\sin{\theta{}}\right)^{\exp{\left[-\frac{\sqrt{\nu t}}{r}\left(1-\frac{R}{r}\right)\right]}}\right.\\[10pt]\times\left.\left(\cos{\theta{}}-\frac{qRe}{2}\, {\frac {\sin^2{\theta }\cos{\theta} \left( {\frac {r}{R}}-1 \right)\frac{{R}^{2}}{{r}^{2}} }{\sqrt {\left( 1-{\frac {{R}^{2}}{{r}^{2}}} \right) ^{2}+4\,{\frac {{R}^{2} \sin^2{\theta}}{{r}^{2}}}+2}}}\right) {\exp{\left[-\left(\frac{r^{2}}{4V_{\infty}^{2}t^{2}}+\frac{qRe\sqrt { \left( 1-{\frac {{R}^{2}}{{r}^{2}}} \right) ^{2}+4\,{\frac {{R}^{2} \sin^2{\theta} }{{r}^{2}}}+2} \left( {\frac {r}{R}}-1 \right)}{8}\right) \right]}}\right)\ln{\left(\dfrac{r}{R}\right)}
				\hspace{3pt}.
	\end{array}	
	\end{equation}
\end{widetext}
Here, attention is also focused on $\ln{(r/R)}$ which continues to enforce both the no-penetration and partial no-slip conditions on these velocity components on the surface as it does in Eq.~\ref{PsiVtVel}. However, the $\ln{(r/R)}$-independent block of terms in ${V}_{\theta_{ss}}$ becomes entirely zero on the wall due to the leading $\left(1-\frac{R^2}{r^2}\right)$ term.

The velocity field of $\psi_p$ is described in Eq.~\ref{CylGravityVelPotVel} which is a function of the velocity components of the superposition of $\psi_{inv}$,~$\psi_{vt}$, and $\psi_{ss}$. This intermediate superposition has already been shown to produce zero velocity at the wall. In addition, $\mu_g$ is a minute number for practical purposes of cylinder flow \cite{Amoloye2024}. Thus, Eq.~\ref{CylGravityVelPotVel} also satisfies the wall conditions.
\begin{widetext}
\begin{equation}
	\label{CylGravityVelPotVel}
    \begin{array}{l}
	V_{r_p}= -\dfrac{2\pi{\mu}^2_{g}}{r^2}\dfrac{\partial^2}{\partial^2 \theta}\left(\left[\dfrac{2{\mu}_{g}}{r}-\begin{array}{l}\left(V_{r_{inv}}+V_{r_{vt}}+V_{r_{ss}}\right)^2 -\left(V_{\theta_{inv}}+V_{\theta_{vt}}+V_{\theta_{ss}}\right)^2\end{array}\right]^{-3/2}\right)\\[10pt]
   V_{\theta_p}= -\dfrac{2\pi{\mu}^2_{g}}{r^2}\dfrac{\partial}{\partial \theta}\left(\left[\dfrac{2{\mu}_{g}}{r}-\begin{array}{l} \left(V_{r_{inv}}+V_{r_{vt}}+V_{r_{ss}}\right)^2 -\left(V_{\theta_{inv}}+V_{\theta_{vt}}+V_{\theta_{ss}}\right)^2\end{array}\right]^{-3/2}\right)\\[10pt]
   +\dfrac{2\pi{\mu}^2_{g}}{r}\dfrac{\partial}{\partial \theta}\left(-\dfrac{3}{2}\left[\dfrac{2{\mu}_{g}}{r}-\begin{array}{l}\left(V_{r_{inv}}+V_{r_{vt}}+V_{r_{ss}}\right)^2 -\left(V_{\theta_{inv}}+V_{\theta_{vt}}+V_{\theta_{ss}}\right)^2\end{array}\right]^{-5/2}\right.\\\left.
   \times \dfrac{\partial}{\partial r}\left(\left[\dfrac{2{\mu}_{g}}{r}-\begin{array}{l}\left(V_{r_{inv}}+V_{r_{vt}}+V_{r_{ss}}\right)^2 -\left(V_{\theta_{inv}}+V_{\theta_{vt}}+V_{\theta_{ss}}\right)^2\end{array}\right]^{-3/2}\right)\right)
	\hspace{3pt},
    \end{array}	
\end{equation}
\end{widetext}
$\psi_{vtsd}$ is modeled on $\psi_{ss}$. Consequently, its velocity field in Eq.~\ref{PsiVtsdVel} satisfy the wall conditions in the same way as in Eq.~\ref{PsiSSVel}.
\begin{widetext}
\begin{equation}
	\label{PsiVtsdVel}
	\begin{array}{l}
	{V}_{r_{vtsd}}=\dfrac{\partial}{\partial \theta}\left(\dfrac{V_{\infty{}}}{5}\left(1-\frac{R^2}{r^2}\right)\left(\cos{\theta{}}-\frac{130}{2Re}\, {\frac {\sin^2{\theta }\cos{\theta} \left( {\frac {
								r}{R}}-1 \right)\frac{{R}^{2}}{{r}^{2}} }{\sqrt {
							\left( 1-{\frac {{R}^{2}}{{r}^{2}}} \right) ^{2}+4\,{\frac {{R}^{2}
									\sin^2{\theta}}{{r}^{2}}}+2}}}\right)\left(\left(\theta{}+\sigma{}\right)\left(\chi\sin{\theta{}}\right)\right)^{\exp{\left[-\frac{V_{\infty}t}{r}\left(1-\frac{R}{r}\right)\right]}}\right.\\[10pt]\left.
				\times {\exp{\left[-\left(\frac{125}{Re}+\frac{r^{2}}{4V_{\infty}^{2}t^{2}}\right)-\left({\exp{(-\frac{130}{Re})}}\left(\frac{r}{\sqrt{R^{2}+4\nu t}}-1\right)^2+\frac{\frac{130}{Re} \sqrt { \left( 1-{\frac {{R}^{2}}{{r}^{2
								}}} \right) ^{2}+4\,{\frac {{R}^{2} \sin^2{\theta} }{{r}^{2}}}+2} \left( {\frac {r(t)}{R}}-1 \right)}{8}\right)\left(\sin^2{\theta{}}+\eta^{\frac{{\exp{\left[-\frac{r^2}{4V_{\infty}^{2}t^{2}}\right]}}}{2}}\right) \right]}}\ln{\left(\dfrac{r}{R}\right)}\right)\\
    {V}_{\theta_{vtsd}}=-\dfrac{V_{\infty{}}}{5}r\left(1-\frac{R^2}{r^2}\right)\left(\cos{\theta{}}-\frac{130}{2Re}\, {\frac {\sin^2{\theta }\cos{\theta} \left( {\frac {
								r}{R}}-1 \right)\frac{{R}^{2}}{{r}^{2}} }{\sqrt {
							\left( 1-{\frac {{R}^{2}}{{r}^{2}}} \right) ^{2}+4\,{\frac {{R}^{2}
									\sin^2{\theta}}{{r}^{2}}}+2}}}\right)\left(\left(\theta{}+\sigma{}\right)\left(\chi\sin{\theta{}}\right)\right)^{\exp{\left[-\frac{V_{\infty}t}{r}\left(1-\frac{R}{r}\right)\right]}}\\[10pt]
				\times {\exp{\left[-\left(\frac{125}{Re}+\frac{r^{2}}{4V_{\infty}^{2}t^{2}}\right)-\left({\exp{(-\frac{130}{Re})}}\left(\frac{r}{\sqrt{R^{2}+4\nu t}}-1\right)^2+\frac{\frac{130}{Re} \sqrt { \left( 1-{\frac {{R}^{2}}{{r}^{2
								}}} \right) ^{2}+4\,{\frac {{R}^{2} \sin^2{\theta} }{{r}^{2}}}+2} \left( {\frac {r(t)}{R}}-1 \right)}{8}\right)\left(\sin^2{\theta{}}+\eta^{\frac{{\exp{\left[-\frac{r^2}{4V_{\infty}^{2}t^{2}}\right]}}}{2}}\right) \right]}}\\[10pt]

                    -\dfrac{\partial}{\partial r}\left(\dfrac{V_{\infty{}}}{5}r\left(1-\frac{R^2}{r^2}\right)\left(\cos{\theta{}}-\frac{130}{2Re}\, {\frac {\sin^2{\theta }\cos{\theta} \left( {\frac {
								r}{R}}-1 \right)\frac{{R}^{2}}{{r}^{2}} }{\sqrt {
							\left( 1-{\frac {{R}^{2}}{{r}^{2}}} \right) ^{2}+4\,{\frac {{R}^{2}
									\sin^2{\theta}}{{r}^{2}}}+2}}}\right)\left(\left(\theta{}+\sigma{}\right)\left(\chi\sin{\theta{}}\right)\right)^{\exp{\left[-\frac{V_{\infty}t}{r}\left(1-\frac{R}{r}\right)\right]}}\right.\\[10pt]\left.
				\times {\exp{\left[-\left(\frac{125}{Re}+\frac{r^{2}}{4V_{\infty}^{2}t^{2}}\right)-\left({\exp{(-\frac{130}{Re})}}\left(\frac{r}{\sqrt{R^{2}+4\nu t}}-1\right)^2+\frac{\frac{130}{Re} \sqrt { \left( 1-{\frac {{R}^{2}}{{r}^{2
								}}} \right) ^{2}+4\,{\frac {{R}^{2} \sin^2{\theta} }{{r}^{2}}}+2} \left( {\frac {r(t)}{R}}-1 \right)}{8}\right)\left(\sin^2{\theta{}}+\eta^{\frac{{\exp{\left[-\frac{r^2}{4V_{\infty}^{2}t^{2}}\right]}}}{2}}\right) \right]}}\right)\ln{\left(\dfrac{r}{R}\right)}
				\hspace{3pt},
	\end{array}	
	\end{equation}
\end{widetext}
Therefore, the consistently defined radial and tangential velocities of $\psi_{vis}$ whose components are Eqs~\ref{PsiInvVel} to ~\ref{PsiVtsdVel} satisfy the wall conditions. Since the azimuthal velocity is entirely dictated by the tangential velocity (Eq.~\ref{Vkappa}), it also satisfies the no-slip condition at the wall.
\section{Calculation of the Vorticity and Laplacian Expressions in the Principal Cartesian Coordinate Wind Axes}\label{CoMm}
The $\mathbf{\omega}$ components in the principal axes are computed as \cite{Amoloyeetal2026b}
\begin{equation}
	\label{dVxdz}
	\begin{array}{l}
		\dfrac{\partial V_{\tilde{x}}}{\partial z}=
		\dfrac{\partial^2 \tilde{\kappa}}{\partial z\partial x}\\[10pt]=\dfrac{\partial }{\partial z}\left(\dfrac{\partial \tilde{\kappa}}{\partial r}\dfrac{\partial r}{\partial x}+\dfrac{\partial \tilde{\kappa}}{r\partial \theta}\dfrac{r\partial\theta }{\partial x}\right)\\[10pt]
		=\left(\dfrac{\partial^2 \tilde{\kappa}}{\partial r^2}\dfrac{\partial r}{\partial x}+\dfrac{\partial^2 \tilde{\kappa}}{r\partial r\partial \theta}\dfrac{r\partial\theta }{\partial x}\right)\dfrac{\partial r}{\partial z}\\[10pt]=-\left(\dfrac{\partial V_{\theta}}{\partial r}\cos{\theta}+\dfrac{\partial V_r}{\partial r}\sin{\theta}\right)\cos{\varphi}\hspace{3pt},
	\end{array}
\end{equation}
\begin{equation}
	\label{dVydz}
	\begin{array}{l}
		\dfrac{\partial V_{\tilde{y}}}{\partial z}=
		\dfrac{\partial^2 \tilde{\kappa}}{\partial z\partial y}\\[10pt]=\dfrac{\partial }{\partial z}\left(\dfrac{\partial \tilde{\kappa}}{\partial r}\dfrac{\partial r}{\partial y}+\dfrac{\partial \tilde{\kappa}}{r\partial \theta}\dfrac{r\partial\theta }{\partial y}\right)\\[10pt]
		=\left(\dfrac{\partial^2 \tilde{\kappa}}{\partial r^2}\dfrac{\partial r}{\partial y}+\dfrac{\partial^2 \tilde{\kappa}}{r\partial r\partial \theta}\dfrac{r\partial\theta }{\partial y}\right)\dfrac{\partial r}{\partial z}\\[10pt]=-\left(\dfrac{\partial V_{\theta}}{\partial r}\sin{\theta}-\dfrac{\partial V_r}{\partial r}\cos{\theta}\right)\cos{\varphi}\hspace{3pt},
	\end{array}
\end{equation}
\begin{equation}
	\label{dVzdx}
	\begin{array}{l}
		\dfrac{\partial V_{z}}{\partial x}=
		\dfrac{\partial^2 \tilde{\kappa}}{\partial x\partial z}\\[10pt]=\dfrac{\partial }{\partial x}\left(\dfrac{\partial \tilde{\kappa}}{\partial r}\dfrac{\partial r}{\partial z}\right)\\[10pt]=\dfrac{\partial^2 \tilde{\kappa}}{\partial r^2}\dfrac{\partial r}{\partial x}\dfrac{\partial r}{\partial z}+\dfrac{\partial^2 \tilde{\kappa}}{r\partial \theta \partial r}\dfrac{r\partial \theta}{\partial x}\dfrac{\partial r}{\partial z}\\[10pt]=-\left(\dfrac{\partial V_{\theta}}{\partial r}\cos{\theta}+\dfrac{\partial V_{r}}{\partial r}\sin{\theta}\right)\cos{\varphi}\hspace{3pt},
	\end{array}
\end{equation}
\begin{equation}
	\label{dVzdy}
	\begin{array}{l}
		\dfrac{\partial V_{z}}{\partial y}=
		\dfrac{\partial^2 \tilde{\kappa}}{\partial y\partial z}\\[10pt]=\dfrac{\partial }{\partial y}\left(\dfrac{\partial \tilde{\kappa}}{\partial r}\dfrac{\partial r}{\partial z}\right)\\[10pt]=\dfrac{\partial^2 \tilde{\kappa}}{\partial r^2}\dfrac{\partial r}{\partial y}\dfrac{\partial r}{\partial z}+\dfrac{\partial^2 \tilde{\kappa}}{r\partial \theta \partial r}\dfrac{r\partial \theta}{\partial y}\dfrac{\partial r}{\partial z}\\[10pt]=-\left(\dfrac{\partial V_{\theta}}{\partial r}\sin{\theta}-\dfrac{\partial V_{r}}{\partial r}\cos{\theta}\right)\cos{\varphi}\hspace{3pt},
	\end{array}
\end{equation}
\begin{widetext}
\begin{equation}
	\label{dVxdy}
	\begin{array}{l}
		\dfrac{\partial V_{\tilde{x}}}{\partial y}=\dfrac{\partial^2 \tilde{\kappa}}{\partial y \partial x}=\dfrac{\partial }{\partial y}\left(\dfrac{\partial \tilde{\kappa}}{\partial r}\dfrac{\partial r}{\partial x}+\dfrac{\partial \tilde{\kappa}}{r\partial \theta}\dfrac{r\partial\theta }{\partial x}\right)\\[10pt] 
		=\left(\dfrac{\partial^2 \tilde{\kappa}}{\partial r^2}\dfrac{\partial r}{\partial y}+\dfrac{\partial^2 \tilde{\kappa}}{r\partial r\partial \theta}\dfrac{r\partial\theta }{\partial y}\right)\dfrac{\partial r}{\partial x}+\dfrac{\partial \tilde{\kappa}}{\partial r}\dfrac{\partial^2 r}{\partial y\partial x}+\left(\dfrac{\partial^2 \tilde{\kappa}}{r\partial r\partial \theta}\dfrac{\partial r}{\partial y}+\dfrac{\partial^2 \tilde{\kappa}}{r^2\partial \theta^2}\dfrac{r\partial\theta }{\partial y}\right)\dfrac{r\partial\theta }{\partial x}+\dfrac{\partial \tilde{\kappa}}{r^2\partial \theta}\dfrac{r^2\partial^2\theta}{\partial y\partial x}\\[10pt]
		=\left(-\dfrac{\partial V_{\theta}}{\partial r}\sin{\theta}\cos{\theta}+\dfrac{\partial V_r}{\partial r}\cos^2{\theta}+\dfrac{V_{\theta}}{r}\sin{\theta}\cos{\theta}\right)-\dfrac{\partial V_r}{\partial r}\sin^2{\theta}-\dfrac{V_r}{r}(1-2\cos^2{\theta})-\dfrac{\partial V_{r}}{r\partial \theta}\sin{\theta}\cos{\theta} \hspace{3pt}, 
	\end{array}
\end{equation}
\begin{equation}
	\label{dVydx}
	\begin{array}{l}
		\dfrac{\partial V_{\tilde{y}}}{\partial x}=\dfrac{\partial^2 \tilde{\kappa}}{\partial x \partial y}=\dfrac{\partial }{\partial x}\left(\dfrac{\partial \tilde{\kappa}}{\partial r}\dfrac{\partial r}{\partial y}+\dfrac{\partial \tilde{\kappa}}{r\partial \theta}\dfrac{r\partial\theta }{\partial y}\right)\\[10pt]
		=\left(\dfrac{\partial^2 \tilde{\kappa}}{\partial r^2}\dfrac{\partial r}{\partial x}+\dfrac{\partial^2 \tilde{\kappa}}{r\partial r\partial \theta}\dfrac{r\partial\theta }{\partial x}\right)\dfrac{\partial r}{\partial y}+\dfrac{\partial \tilde{\kappa}}{\partial r}\dfrac{\partial^2 r}{\partial x\partial y}+\left(\dfrac{\partial^2 \tilde{\kappa}}{r\partial r\partial \theta}\dfrac{\partial r}{\partial x}+\dfrac{\partial^2 \tilde{\kappa}}{r^2\partial \theta^2}\dfrac{r\partial\theta }{\partial x}\right)\dfrac{r\partial\theta }{\partial y}+\dfrac{\partial \tilde{\kappa}}{r^2\partial \theta}\dfrac{r^2\partial^2\theta}{\partial x\partial y}\\[10pt]
		=\left(-\dfrac{\partial V_{\theta}}{\partial r}\sin{\theta}\cos{\theta}-\dfrac{\partial V_r}{\partial r}\sin^2{\theta}+\dfrac{V_{\theta}}{r}\sin{\theta}\cos{\theta}\right)+\dfrac{\partial V_r}{\partial r}\cos^2{\theta}-\dfrac{V_r}{r}(1-2\cos^2{\theta})-\dfrac{\partial V_{r}}{r\partial \theta}\sin{\theta}\cos{\theta}\hspace{3pt},
	\end{array}
\end{equation}
\begin{equation}
	\label{omegax}
	\begin{array}{l}
		\omega_{x}=\dfrac{\partial V_{z}}{\partial y}-\dfrac{\partial V_{\tilde{y}}}{\partial z}\\[10pt]=
		-\left(\dfrac{\partial V_{\theta}}{\partial r}\sin{\theta}-\dfrac{\partial V_{r}}{\partial r}\cos{\theta}\right)\cos{\varphi}-\left(-\left(\dfrac{\partial V_{\theta}}{\partial r}\sin{\theta}-\dfrac{\partial V_r}{\partial r}\cos{\theta}\right)\cos{\varphi}\right)=0 \hspace{3pt},
	\end{array}
\end{equation}
\begin{equation}
	\label{omegay}
	\begin{array}{l}
		\omega_{y}=\dfrac{\partial V_{\tilde{x}}}{\partial z}-\dfrac{\partial V_{z}}{\partial x}
		\\[10pt]=-\left(\dfrac{\partial V_{\theta}}{\partial r}\cos{\theta}+\dfrac{\partial V_r}{\partial r}\sin{\theta}\right)\cos{\varphi}-\left(-\left(\dfrac{\partial V_{\theta}}{\partial r}\cos{\theta}+\dfrac{\partial V_{r}}{\partial r}\sin{\theta}\right)\cos{\varphi}\right)=0 \hspace{3pt}
	\end{array}
\end{equation}
\end{widetext}
and
\begin{widetext}
\begin{equation}
	\label{omegaztilde}
	\begin{array}{l}
		\omega_{z}=\dfrac{\partial V_{\tilde{y}}}{\partial x}-\dfrac{\partial V_{\tilde{x}}}{\partial y}\\[10pt]
		=\left(-\dfrac{\partial V_{\theta}}{\partial r}\sin{\theta}\cos{\theta}-\dfrac{\partial V_r}{\partial r}\sin^2{\theta}+\dfrac{V_{\theta}}{r}\sin{\theta}\cos{\theta}\right)+\dfrac{\partial V_r}{\partial r}\cos^2{\theta}-\dfrac{V_r}{r}(1-2\cos^2{\theta})-\dfrac{\partial V_{r}}{r\partial \theta}\sin{\theta}\cos{\theta}\\[10pt]
		-\left[\left(-\dfrac{\partial V_{\theta}}{\partial r}\sin{\theta}\cos{\theta}+\dfrac{\partial V_r}{\partial r}\cos^2{\theta}+\dfrac{V_{\theta}}{r}\sin{\theta}\cos{\theta}\right)-\dfrac{\partial V_r}{\partial r}\sin^2{\theta}-\dfrac{V_r}{r}(1-2\cos^2{\theta})-\dfrac{\partial V_{r}}{r\partial \theta}\sin{\theta}\cos{\theta}\right]\\[10pt]
		=0\hspace{3pt}.
	\end{array}
\end{equation}
\end{widetext}

The pressure does not change with the rotation of the Cartesian axes \cite{Schlichting1979,Amoloye2020,Amoloye2024,Amoloyeetal2026b}. The Laplacian of $\tilde{\kappa}$ needed to obtain the viscous term of the pressure field (Eq.~\ref{Velpol}) is calculated as\cite{Amoloye2020,Amoloyeetal2026b}
\begin{widetext}
		\begin{equation}
			\label{dVxdx}
			\begin{array}{l}
				\dfrac{\partial V_{\tilde{x}}}{\partial x}=\dfrac{\partial^2 \tilde{\kappa}}{\partial x^2}=\dfrac{\partial }{\partial x}\left(\dfrac{\partial \tilde{\kappa}}{\partial r}\dfrac{\partial r}{\partial x}+\dfrac{\partial \tilde{\kappa}}{r\partial \theta}\dfrac{r\partial\theta }{\partial x}\right)\\[10pt]
				=\left(\dfrac{\partial^2 \tilde{\kappa}}{\partial r^2}\dfrac{\partial r}{\partial x}+\dfrac{\partial^2 \tilde{\kappa}}{r\partial r\partial \theta}\dfrac{r\partial\theta }{\partial x}\right)\dfrac{\partial r}{\partial x}+\dfrac{\partial \tilde{\kappa}}{\partial r}\dfrac{\partial^2 r}{\partial x^2}+\left(\dfrac{\partial^2 \tilde{\kappa}}{r\partial r\partial \theta}\dfrac{\partial r}{\partial x}+\dfrac{\partial^2 \tilde{\kappa}}{r^2\partial \theta^2}\dfrac{r\partial\theta }{\partial x}\right)\dfrac{r\partial\theta }{\partial x}+\dfrac{\partial \tilde{\kappa}}{r^2\partial \theta}\dfrac{r^2\partial^2\theta}{\partial x^2}\\[10pt]
				=-\left(\dfrac{\partial V_{\theta}}{\partial r}\cos^2{\theta}+\dfrac{\partial V_r}{\partial r}\sin{\theta}\cos{\theta}+\dfrac{V_{\theta}}{r}(1-\cos^2{\theta})\right)-\left(\dfrac{\partial V_r}{\partial r}-\dfrac{V_r}{r}\right)\sin{\theta}\cos{\theta} +\dfrac{\partial V_{r}}{r\partial \theta}\sin^2{\theta}\hspace{3pt},
			\end{array}
		\end{equation}
		\begin{equation}
			\label{dVydy}
			\begin{array}{l}
				\dfrac{\partial V_{\tilde{y}}}{\partial y}=\dfrac{\partial^2 \tilde{\kappa}}{\partial y^2}=\dfrac{\partial }{\partial y}\left(\dfrac{\partial \tilde{\kappa}}{\partial r}\dfrac{\partial r}{\partial y}+\dfrac{\partial \tilde{\kappa}}{r\partial \theta}\dfrac{r\partial\theta }{\partial y}\right)\\[10pt]
				=\left(\dfrac{\partial^2 \tilde{\kappa}}{\partial r^2}\dfrac{\partial r}{\partial y}+\dfrac{\partial^2 \tilde{\kappa}}{r\partial r\partial \theta}\dfrac{r\partial\theta }{\partial y}\right)\dfrac{\partial r}{\partial y}+\dfrac{\partial \tilde{\kappa}}{\partial r}\dfrac{\partial^2 r}{\partial y^2}+\left(\dfrac{\partial^2 \tilde{\kappa}}{r\partial r\partial \theta}\dfrac{\partial r}{\partial y}+\dfrac{\partial^2 \tilde{\kappa}}{r^2\partial \theta^2}\dfrac{r\partial\theta }{\partial y}\right)\dfrac{r\partial\theta }{\partial y}+\dfrac{\partial \tilde{\kappa}}{r^2\partial \theta}\dfrac{r^2\partial^2\theta}{\partial y^2}\\[10pt]
				=-\left(\dfrac{\partial V_{\theta}}{\partial r}\sin^2{\theta}-\dfrac{\partial V_r}{\partial r}\sin{\theta}\cos{\theta}+\dfrac{V_{\theta}}{r}(1-\sin^2{\theta})\right)+\left(\dfrac{\partial V_r}{\partial r}-\dfrac{V_r}{r}\right)\sin{\theta}\cos{\theta}+\dfrac{\partial V_{r}}{r\partial \theta}\cos^2{\theta}\hspace{3pt},
			\end{array}
		\end{equation}
		\begin{equation}
			\label{dVzdz}
			\begin{array}{l}
				\dfrac{\partial V_{z}}{\partial z}=\dfrac{\partial^2 \tilde{\kappa}}{\partial z^2}=\dfrac{\partial }{\partial z}\left(\dfrac{\partial \tilde{\kappa}}{\partial r}\dfrac{\partial r}{\partial z}\right)\\[10pt]=\dfrac{\partial^2 \tilde{\kappa}}{\partial r^2}\dfrac{\partial r}{\partial z}\dfrac{\partial r}{\partial z}+\dfrac{\partial \tilde{\kappa}}{\partial r}\dfrac{\partial^2 r}{\partial z^2}\\[10pt]=-\left(\dfrac{\partial V_{\theta}}{\partial r}\cos^2{\varphi}+\dfrac{V_{\theta}}{r}\sin^2{\varphi}\right)\hspace{3pt},
			\end{array}
		\end{equation}
        \end{widetext}
		and       
\begin{widetext}
		\begin{equation}
			\label{crossderive}
			\begin{array}{l}
				{{\nabla{}}}^2\tilde{\kappa}=\dfrac{\partial V_{\tilde{x}}}{\partial x}+\dfrac{\partial V_{\tilde{y}}}{\partial y}+\dfrac{\partial V_{z}}{\partial z}=\dfrac{\partial^2 \tilde{\kappa}}{\partial x^2}+\dfrac{\partial^2 \tilde{\kappa}}{\partial y^2}+\dfrac{\partial^2 \tilde{\kappa}}{\partial z^2}\\[10pt]
				=-\left(\dfrac{\partial V_{\theta}}{\partial r}\cos^2{\theta}+\dfrac{\partial V_r}{\partial r}\sin{\theta}\cos{\theta}+\dfrac{V_{\theta}}{r}(1-\cos^2{\theta})\right)-\left(\dfrac{\partial V_r}{\partial r}-\dfrac{V_r}{r}\right)\sin{\theta}\cos{\theta}\\[10pt]+\dfrac{\partial V_{r}}{r\partial \theta}\sin^2{\theta}
				-\left(\dfrac{\partial V_{\theta}}{\partial r}\sin^2{\theta}-\dfrac{\partial V_r}{\partial r}\sin{\theta}\cos{\theta}+\dfrac{V_{\theta}}{r}(1-\sin^2{\theta})\right)\\[10pt]+\left(\dfrac{\partial V_r}{\partial r}-\dfrac{V_r}{r}\right)\sin{\theta}\cos{\theta}+\dfrac{\partial V_{r}}{r\partial \theta}\cos^2{\theta}
				-\left(\dfrac{\partial V_{\theta}}{\partial r}\cos^2{\varphi}+\dfrac{V_{\theta}}{r}\sin^2{\varphi}\right)\\[10pt]
				=-\dfrac{\partial V_{\theta}}{\partial r}-\dfrac{V_{\theta}}{r}+\dfrac{\partial V_{r}}{r\partial \theta}
				-\dfrac{\partial V_{\theta}}{\partial r}\cos^2{\varphi}-\dfrac{V_{\theta}}{r}\sin^2{\varphi}\hspace{3pt}.
			\end{array}
		\end{equation}
\end{widetext}

\color{black}
\section{Verification of the RPT Quasi-potential Simplification of the Navier-Stokes Problem}\label{ResCh}
The numerical verification of the $\tilde{\kappa}$ simplification of the Navier-Stokes problem presented here is carried out on a $64$-bit double precision computer. The machine-epsilon, $\epsilon_{mach}$ is approximately equal to $2.22 \times 10^{-16}$ for this double-precision hardware \cite{Higham2002,IEEE754_2019,Oberkampf2010}. However, the practical scaled round-off floor is between $10^{-15}$ and $10^{-12}$ due to the volume of accumulated arithmetic operations in the evaluation \cite{Higham2002,Oberkampf2010}. 

\begin{figure*}
	\centering
	\subfloat[$\theta=\pi/4$]{\includegraphics[width=0.25\textwidth,trim={0cm 0cm 0cm 0cm},clip]{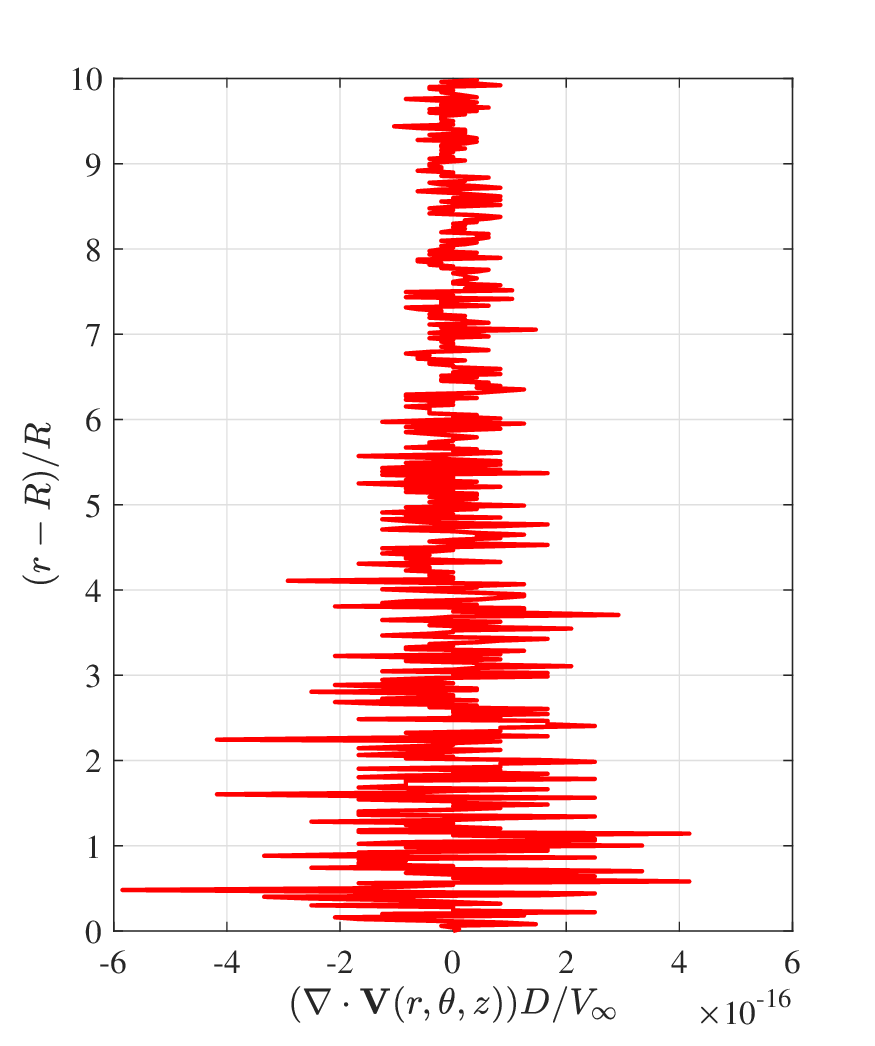}\label{Fig18a}}
	\subfloat[$\theta=-\pi$]{\includegraphics[width=0.25\textwidth,trim={0cm 0cm 0cm 0cm},clip]{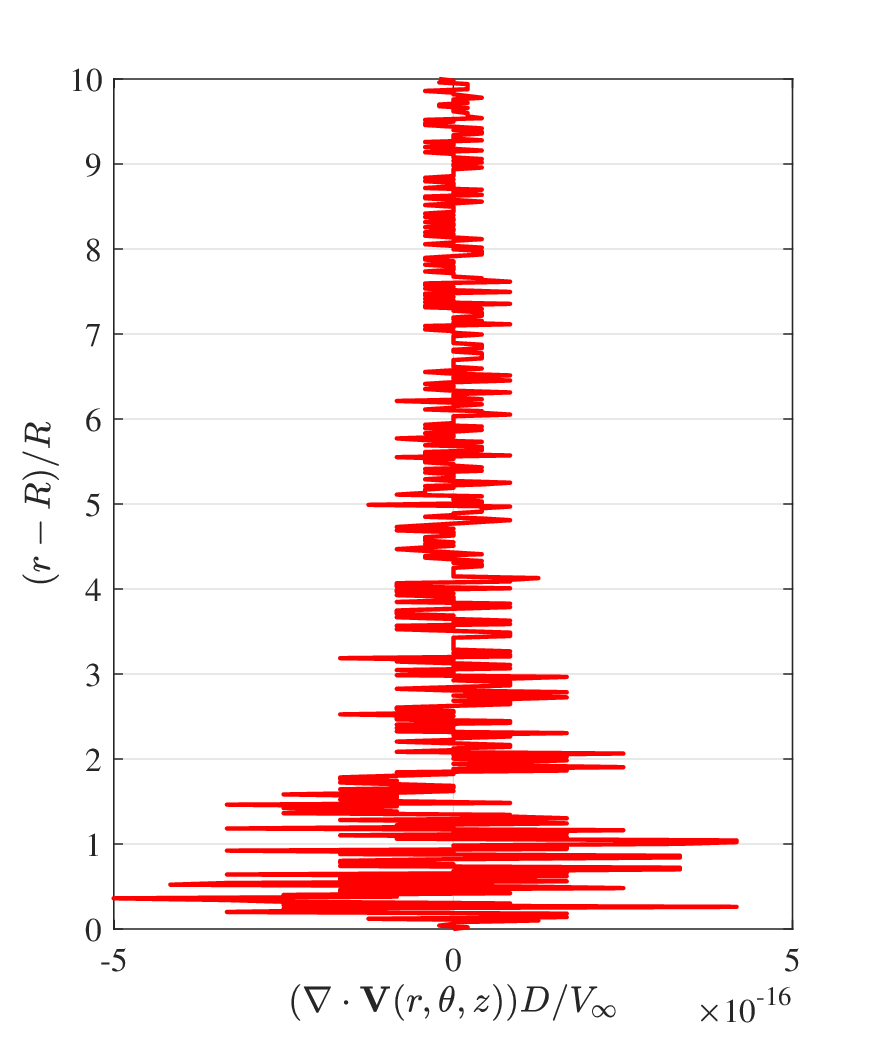}\label{Fig18b}}
	\subfloat[$\theta=-3\pi/4$]{\includegraphics[width=0.25\textwidth,trim={0cm 0cm 0cm 0cm},clip]{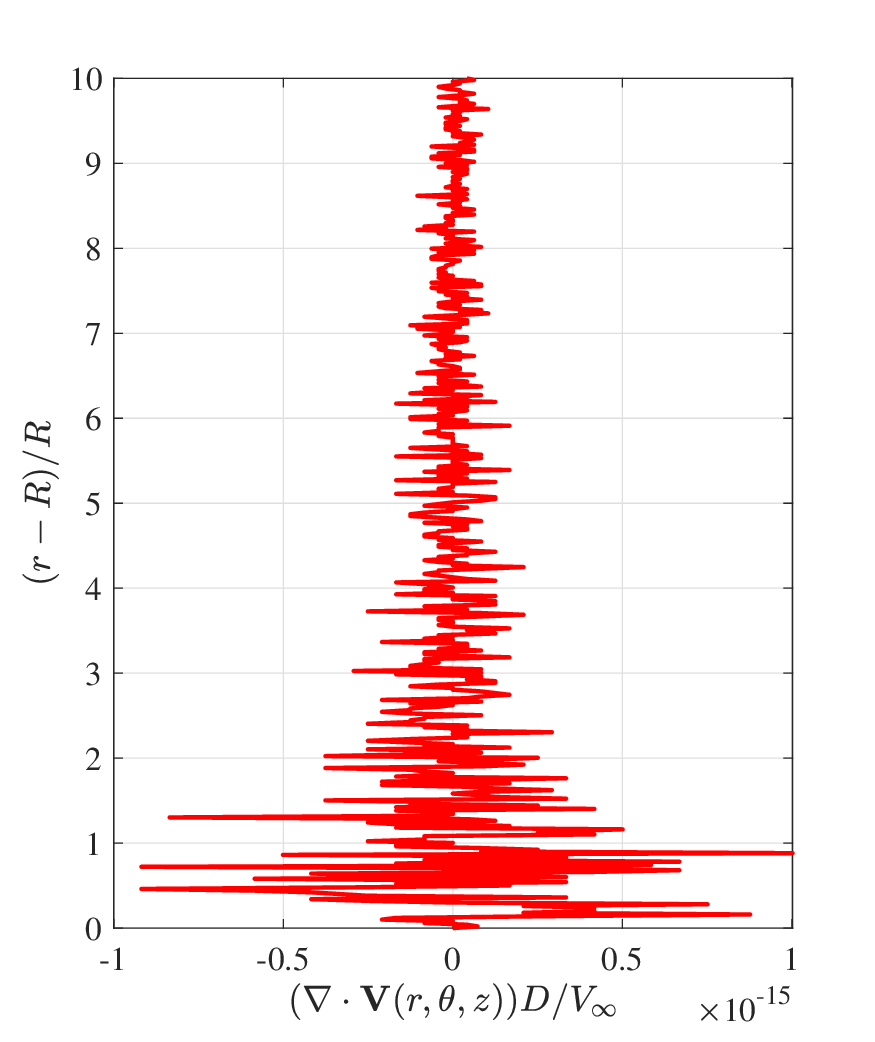}\label{Fig18c}}
	\subfloat[$\theta=-\pi/2$]{\includegraphics[width=0.25\textwidth,trim={0cm 0cm 0cm 0cm},clip]{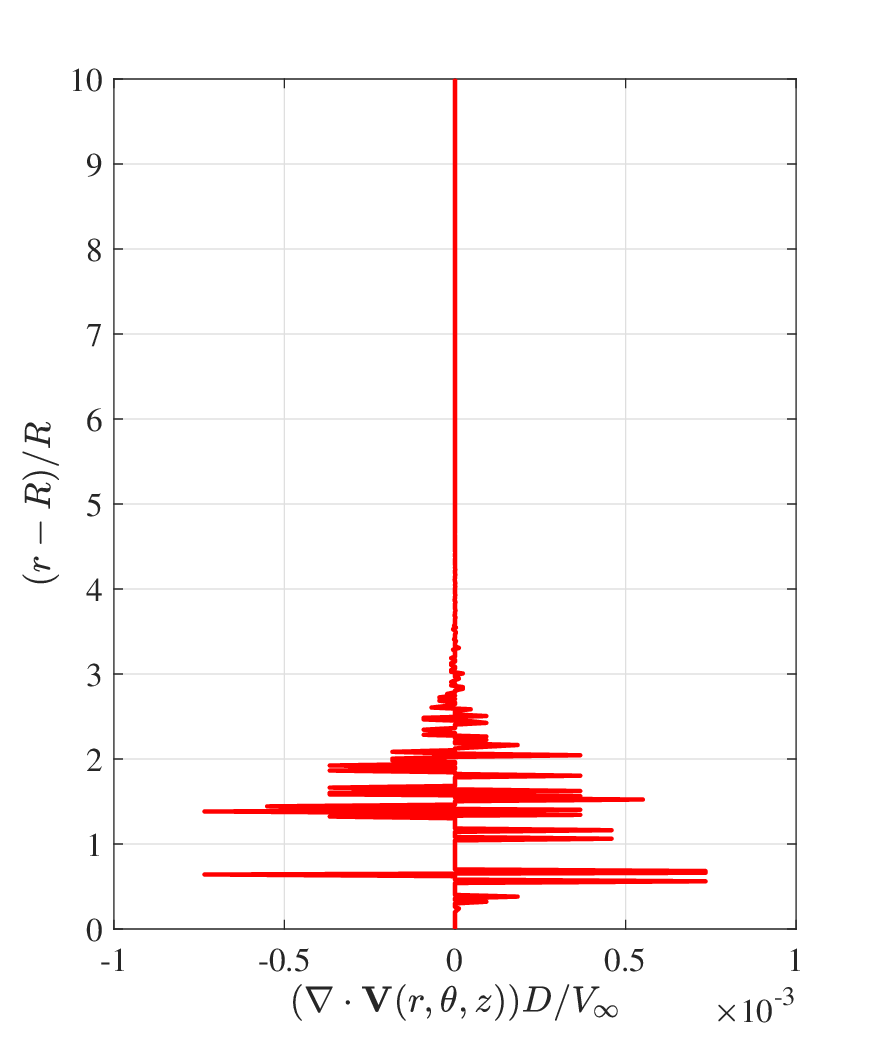}\label{Fig18d}}
	\caption{Continuity equation check on the radial line extending outwards from the $(a)$ windward side, $(b)$ crest, $(c)$ leeward side, and $(d)$ wake center line in the $\varphi=\pi/2$ plane ($z=0$). Flow conditions are the same as in figure~\ref{Fig7}.}
	\label{Fig18}
\end{figure*}

Figure~\ref{Fig18} presents the verification of $R_c$ in the $z=0$ plane. This is observed to be below the round-off floor for the windward face (figure~\ref{Fig18a}, crest (figure~\ref{Fig18b}), and leeward face (figure\ref{Fig18c}) but not the center line of the wake (figure~\ref{Fig18d}). Outlier points experiencing extremely high but finite normalized velocity jumps ($\mathcal{O}(10^{16})$) exist on the cylinder rear axis and around it in RPT. These are points in the center or near the center of the modeled vortices \cite{Amoloye2024}. Differential operations on the velocity field at these outlier points also yield results with extremely high orders of magnitude. The floating-point spacing at these points scales with the magnitude of the variable concerned, and the unit in the last place (ULP) scales as $ULP\approx |variable| \cdot \epsilon_{mach}$ \cite{Goldberg1991}. ULP gives an idea of when machine-level round-off noise occurs. For the normalized velocity in an outlier point, $ULP\approx 1$. For $R_c$, it is higher. Thus, figure~\ref{Fig18d} also verifies $R_c=0$.   

\begin{figure*}
	\centering
	\includegraphics[width=\textwidth,trim={0cm 0cm 0cm 0cm},clip]{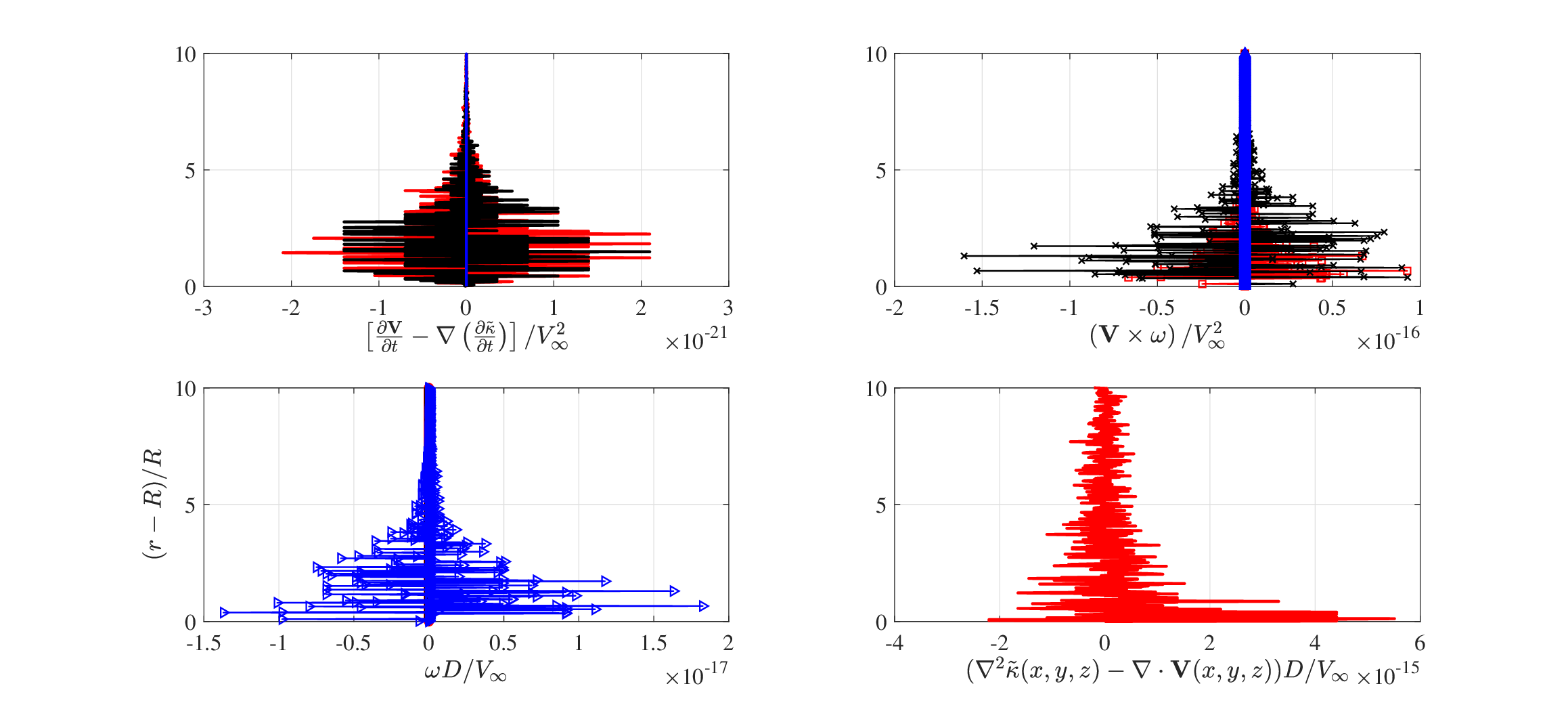}
	\caption{Analysis of the residuals of the pressure field on a radial line extending outwards from the windward side ($\theta=\pi/4$) in the $\varphi=\pi/2$ plane ($z=0$). Flow conditions are the same as in figure~\ref{Fig7}. $(a)$ $\left[\frac{\partial V_{\tilde{x}}}{\partial t}-\frac{\partial}{\partial x}\left(\frac{\partial \tilde{\kappa}}{\partial t}\right)\right]/V^2_{\infty}$ (\color{red}-\color{black}); $\left[\frac{\partial V_{\tilde{y}}}{\partial t}-\frac{\partial}{\partial y}\left(\frac{\partial \tilde{\kappa}}{\partial t}\right)\right]/V^2_{\infty}$ (\color{black}-\color{black}); $\left[\frac{\partial V_z}{\partial t}-\frac{\partial}{\partial z}\left(\frac{\partial \tilde{\kappa}}{\partial t}\right)\right]/V^2_{\infty}$ (\color{blue}-\color{black})  $(b)$ $\left(\mathbf{V} \times \omega \right)_x/V^2_{\infty}$ (\color{red}-$\square$\color{black}); $\left(\mathbf{V} \times \omega \right)_y/V^2_{\infty}$ (\color{black}-$\times$\color{black}); $\left(\mathbf{V} \times \omega \right)_z/V^2_{\infty}$ (\color{blue}-$\triangle$\color{black}) $(c)$ $\omega_xD/V_{\infty}$ (\color{red}-$\circ$\color{black}); $\omega_y D/V_{\infty}$ (\color{black}-$+$\color{black}); $\omega_z D/V_{\infty}$ (\color{blue}-$\triangle$\color{black}), and $(d)$ $(\nabla^2\tilde{\kappa}(x,y,z)-\nabla \cdot \mathbf{V}(x,y,z))D/V_{\infty}$ (\color{red}-\color{black})}\label{Fig19}
\end{figure*}

\begin{figure*}
	\centering
	\includegraphics[width=\textwidth,trim={0cm 0cm 0cm 0cm},clip]{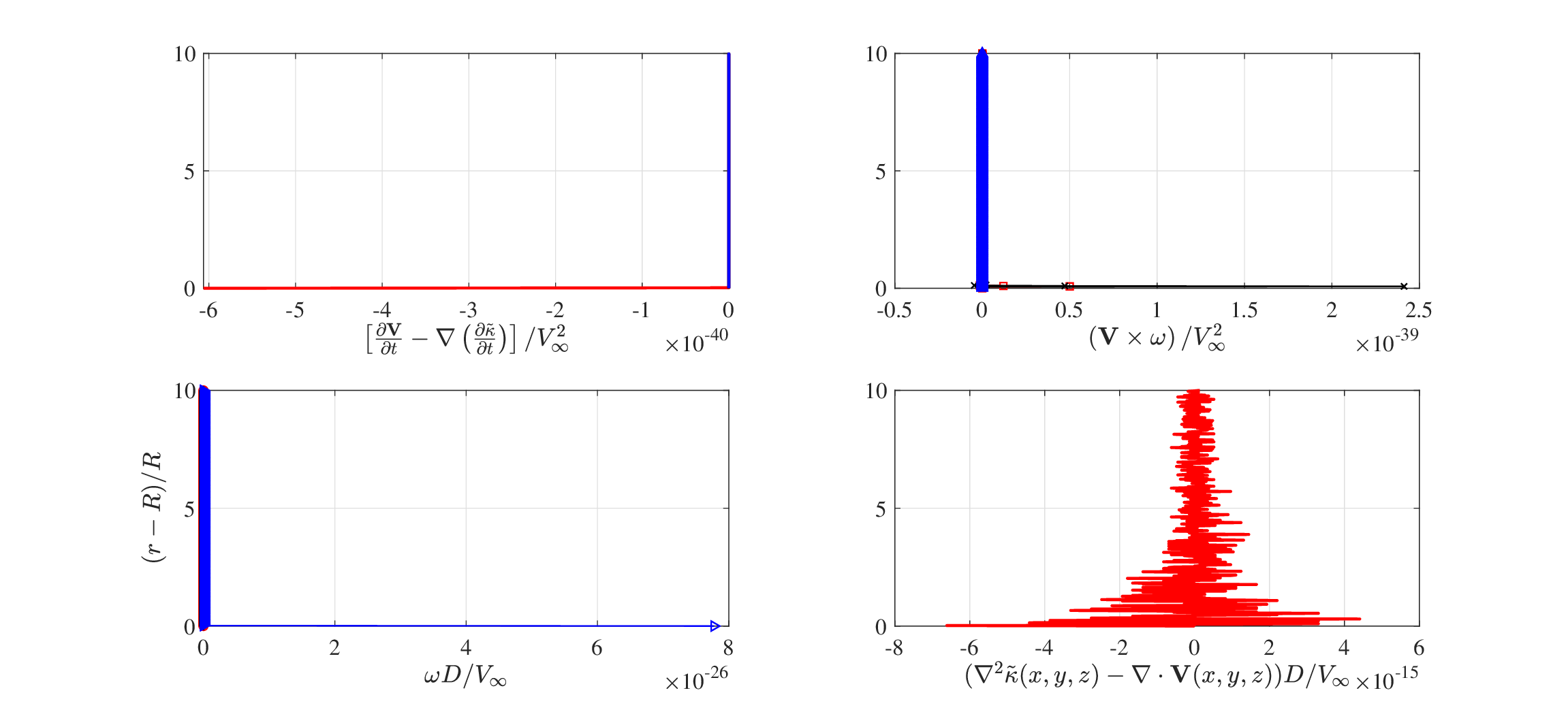}
	\caption{Analysis of the residuals of the pressure field on a radial line extending outwards from the crest ($\theta=-\pi$) in the $\varphi=\pi/2$ plane ($z=0$). Flow conditions are the same as in figure~\ref{Fig7}. See figure~\ref{Fig19} for the legend.}\label{Fig20}
\end{figure*}

\begin{figure*}
	\centering
	\includegraphics[width=\textwidth,trim={0cm 0cm 0cm 0cm},clip]{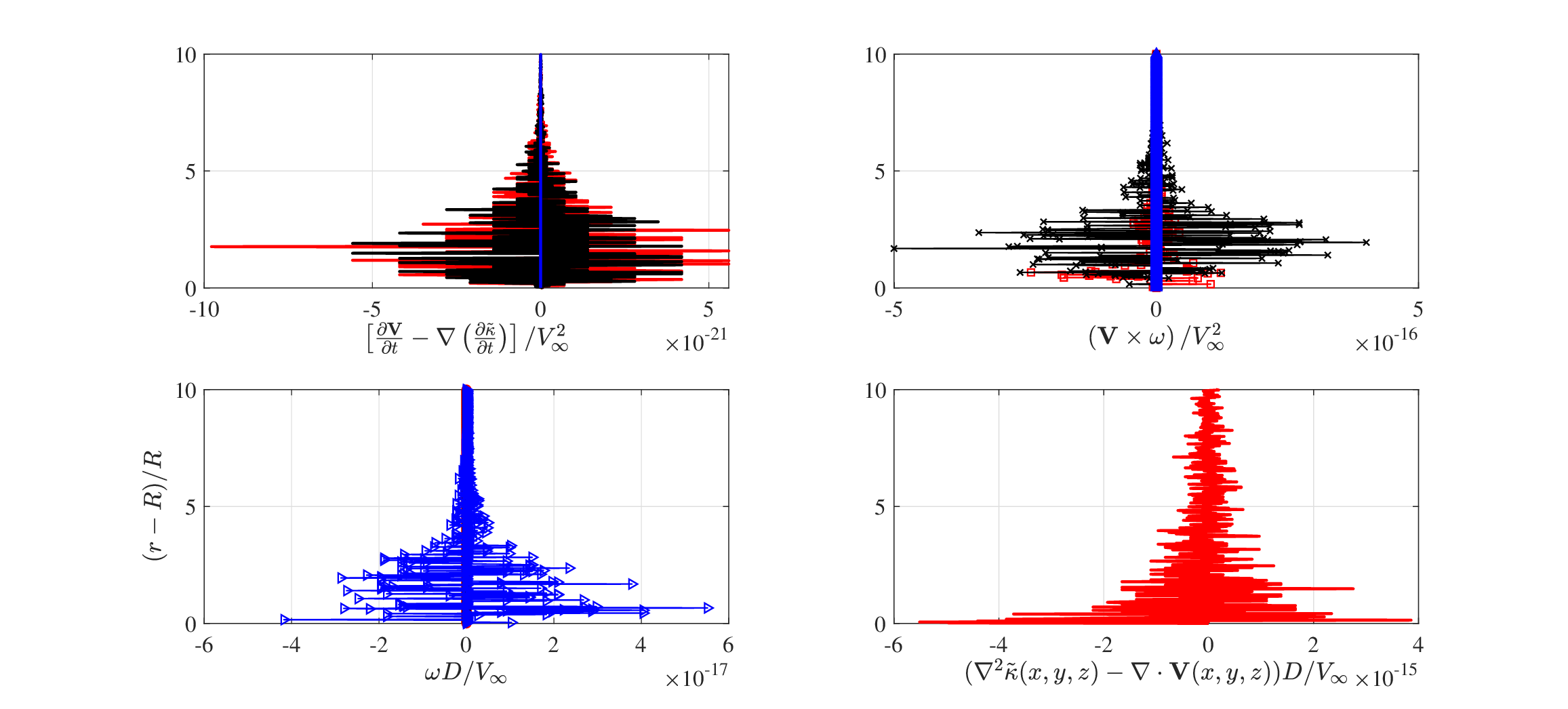}
	\caption{Analysis of the residuals of the pressure field on a radial line extending outwards from the leeward side ($\theta=-3\pi/4$) in the $\varphi=\pi/2$ plane ($z=0$). Flow conditions are the same as in figure~\ref{Fig7}. See figure~\ref{Fig19} for the legend.}\label{Fig21}
\end{figure*}

\begin{figure*}
	\centering
	\includegraphics[width=\textwidth,trim={0cm 0cm 0cm 0cm},clip]{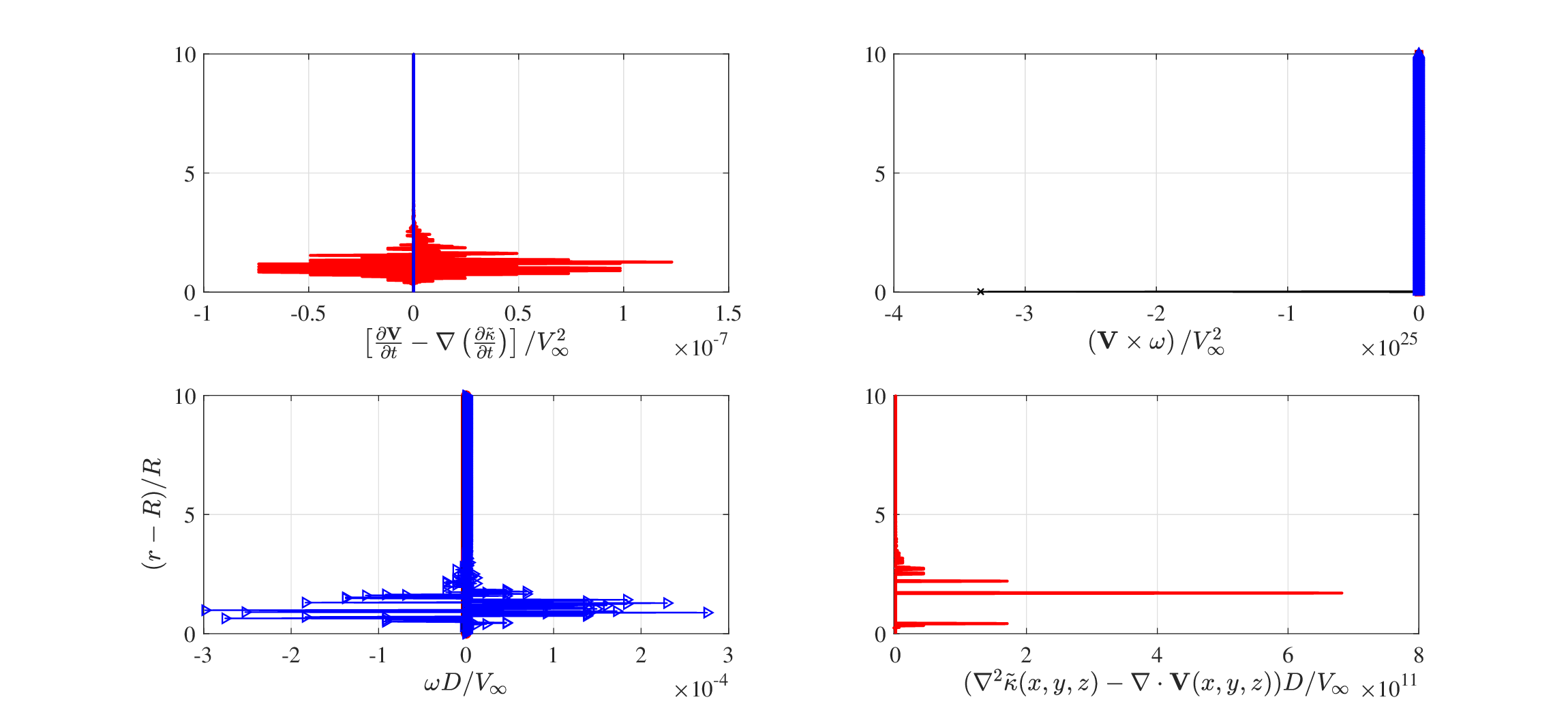}
	\caption{Analysis of the residuals of the pressure field on a radial line extending outwards from the rear stagnation point (wake center line, $\theta=-\pi/2$) in the $\varphi=\pi/2$ plane ($z=0$). Flow conditions are the same as in figure~\ref{Fig7}. See figure~\ref{Fig19} for the legend. }\label{Fig22}
\end{figure*}

Figures~\ref{Fig19}-\ref{Fig22} present the verification analysis of $\mathbf{R}_{m}$ (and $\mathbf{R}_{\omega}$) in the  $z=0$ plane. All the components of $\mathbf{R}_{m}$ as well as the vorticity are consistently below or around the round-off floor for the windward face (figure~\ref{Fig19}), crest (figure~\ref{Fig20}), and leeward face (figure~\ref{Fig21}). Figure~\ref{Fig22} also presents the singular behavior of the outlier points that sometimes exist on the rear axis of the cylinder. The Lamb vector (figure~\ref{Fig22}b) and the Laplacian (figure~\ref{Fig22}d) are the most impacted. The maximum magnitudes of the normalized Lamb vector and the normalized Laplacian are $\mathcal{O}(10^{25})$ and $\mathcal{O}(10^{28})$, respectively. These put their respective ULPs at approximately $10^{9}$ and $10^{12}$. These imply that verification of $\mathbf{R}_{m}$ (and $\mathbf{R}_{\omega}$) fails at some of these outlier points for RPT. 

\begin{figure*}
	\centering
	\subfloat[$\varphi=\pi/4$]{\includegraphics[width=0.25\textwidth,trim={0cm 0cm 0cm 0cm},clip]{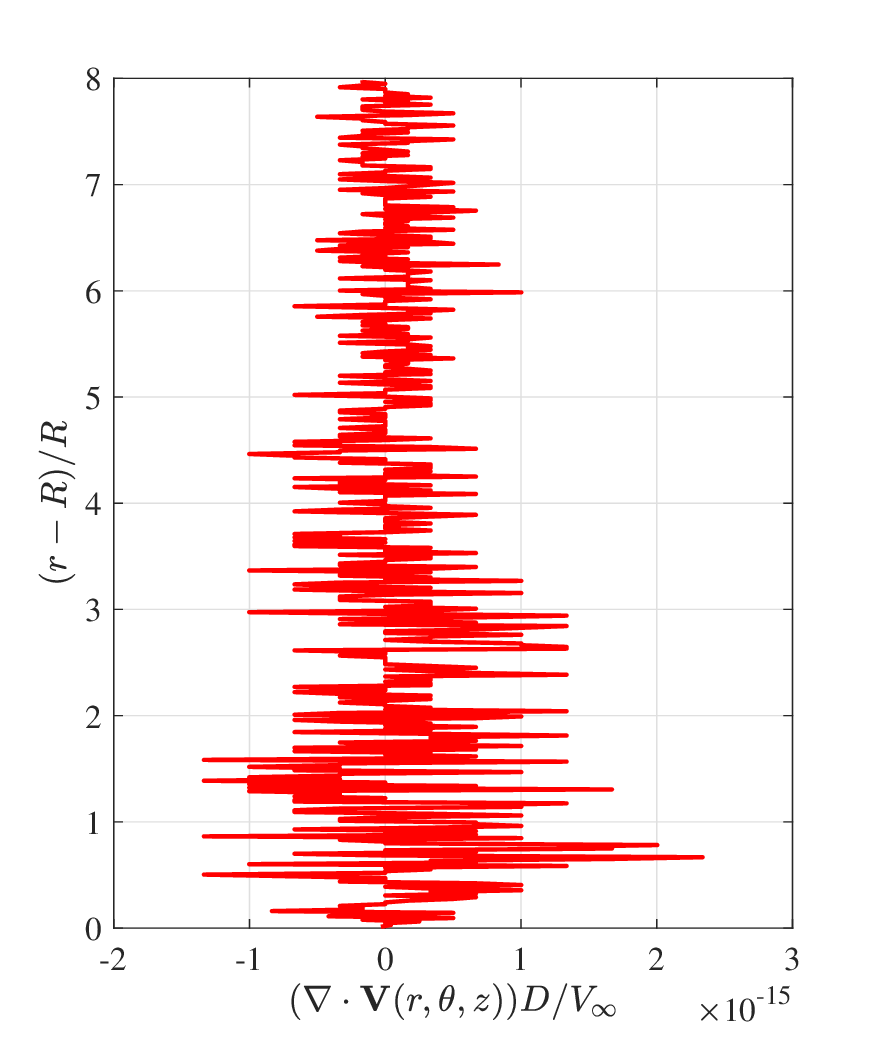}\label{Fig23a}}
	\subfloat[$\varphi=3\pi/4$]{\includegraphics[width=0.25\textwidth,trim={0cm 0cm 0cm 0cm},clip]{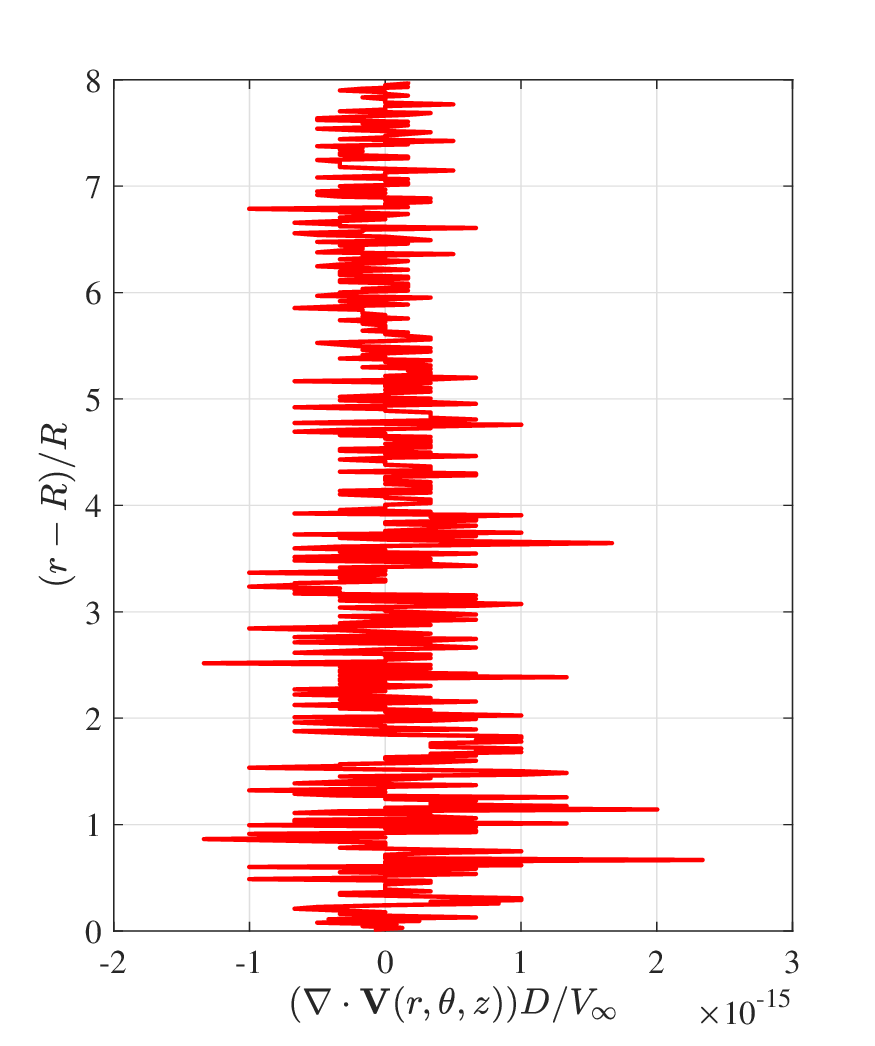}\label{Fig23b}}
		\caption{Continuity equation check on the radial line extending outwards from the crest ($\theta=-\pi$) in the ($a$)$\varphi=\pi/4$ plane ($z>0$) and ($b$)$\varphi=3\pi/4$ plane ($z<0$). Flow conditions are the same as in figure~\ref{Fig7}.}
	\label{Fig23}
\end{figure*}

\begin{figure*}
	\centering
	\includegraphics[width=\textwidth,trim={0cm 0cm 0cm 0cm},clip]{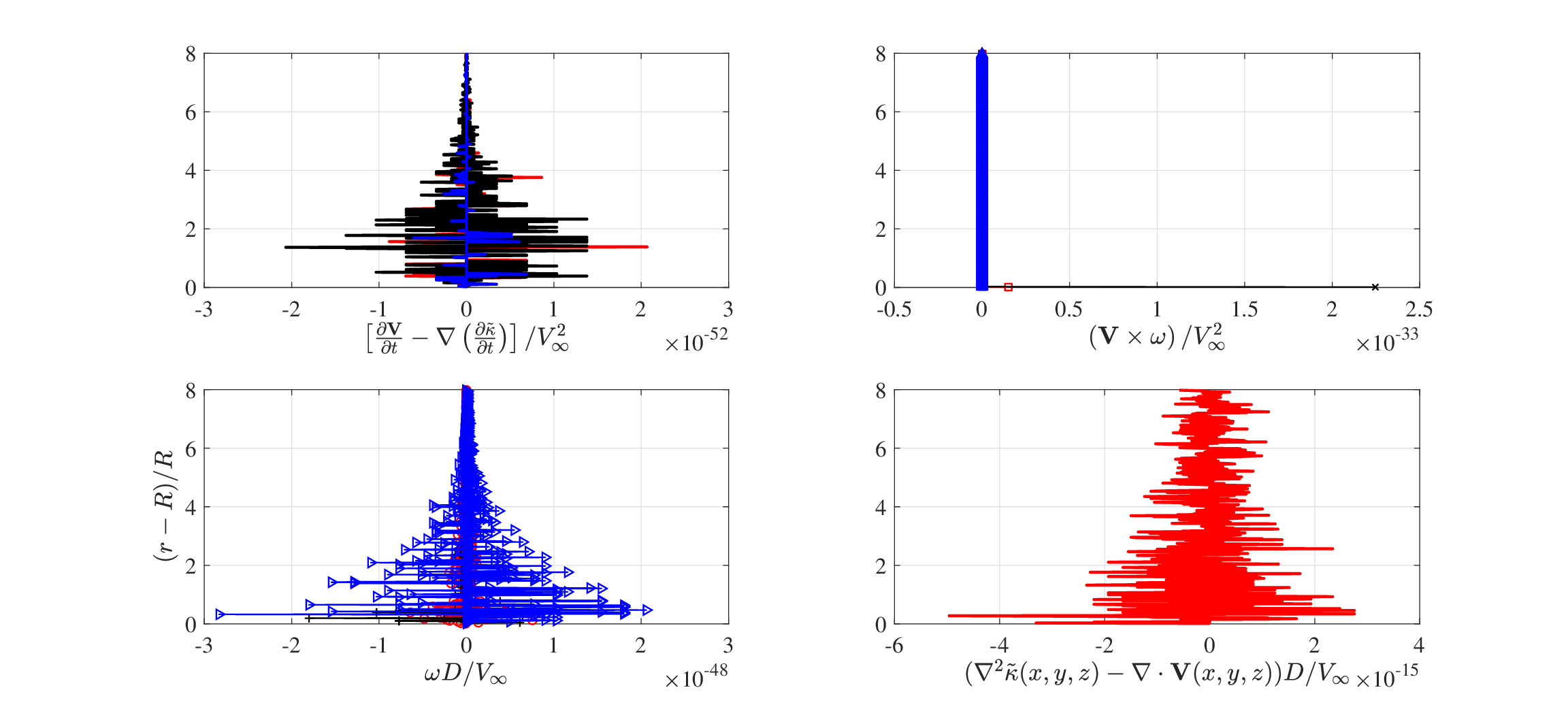}
	\caption{Analysis of the residuals of the pressure field on a radial line extending outwards from the crest ($\theta=-\pi$) in the $\varphi=\pi/4$ plane ($z>0$). Flow conditions are the same as in figure~\ref{Fig7}. See figure~\ref{Fig19} for the legend.}\label{Fig24}
\end{figure*}

\begin{figure*}
	\centering
	\includegraphics[width=\textwidth,trim={0cm 0cm 0cm 0cm},clip]{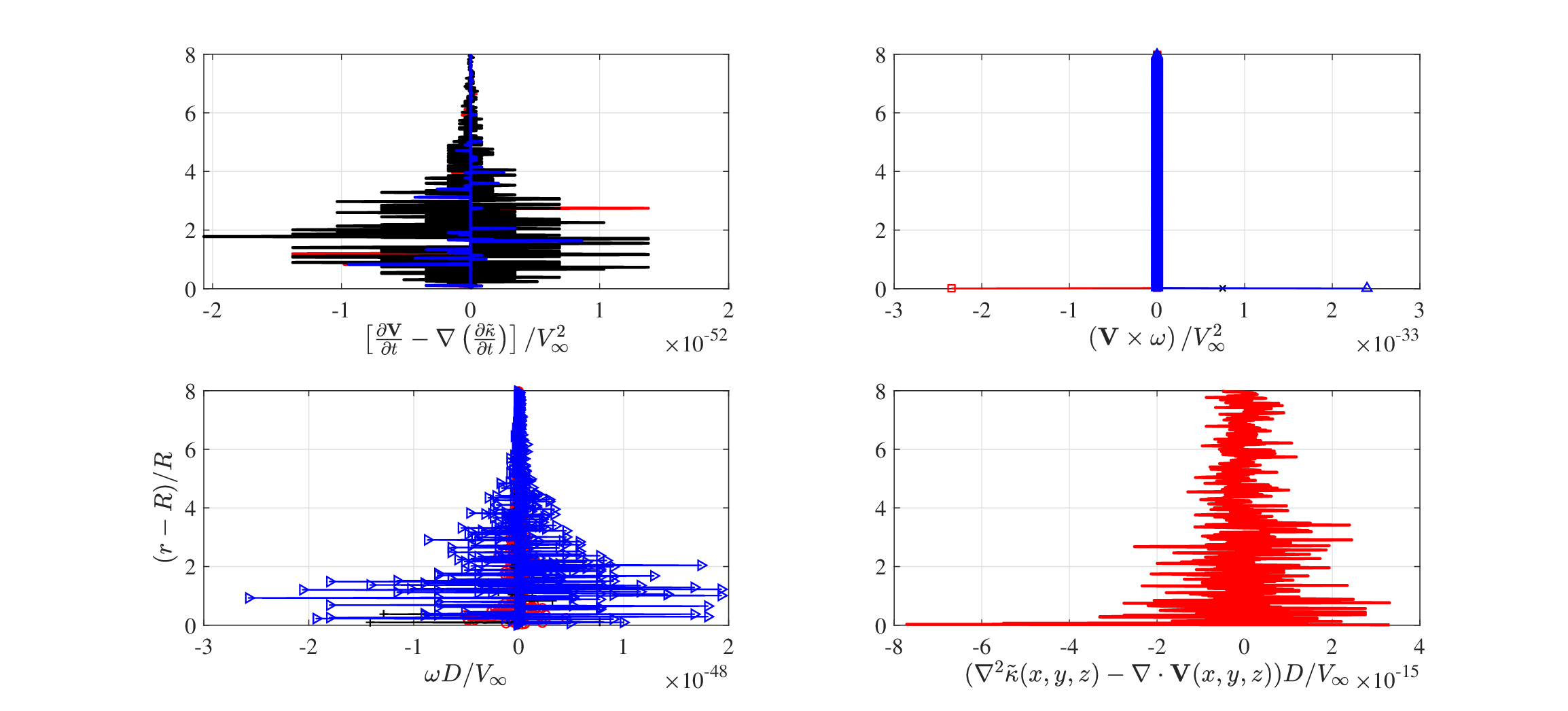}
	\caption{Analysis of the residuals of the pressure field on a radial line extending outwards from the crest ($\theta=-\pi$) in the $\varphi=3\pi/4$ plane ($z<0$). Flow conditions are the same as in figure~\ref{Fig7}. See figure~\ref{Fig19} for the legend.}\label{Fig25}
\end{figure*}

Figures~\ref{Fig23}-~\ref{Fig25} present the verification of $Rc$, $\mathbf{R}_{m}$ and $\mathbf{R}_{\omega}$ at the crest of the cylinder and away from the $z=0$ plane where the flow is three-dimensional. All the concerned parameters are consistently below or around the round-off floor. 

\color{black}
\end{document}